\documentclass[epjc3,twocolumn]{svjour3} 
\pdfoutput=1
\usepackage[utf8]{inputenc}

\usepackage{graphicx}

\usepackage{color} 
\usepackage{amssymb}
\usepackage{amsmath}

\usepackage{verbatim}

\newcommand{\mr}{\mathrm}
\newcommand{\mc}{\mathcal}

\newcommand{\wpwpjj}{W^+W^+jj}

\newcommand{\wpmzjj}{W^\pm Zjj}
\newcommand{\wzjj}{WZjj}
\newcommand{\wpwmjj}{W^+W^-jj}
\newcommand{\zzjj}{ZZjj}
\newcommand{\zjj}{Zjj}

\newcommand{\muf}{\mu_\mr{F}}
\newcommand{\mur}{\mu_\mr{R}}
\newcommand{\xif}{\xi_\mr{F}}
\newcommand{\xir}{\xi_\mr{R}}

\newcommand{\emmvjj}{\nu_e e^+\mu^-\mu^+jj}

\newcommand{\VBFNLO}{{\tt{VBFNLO}}}
\newcommand{\MOCANLORECOLA}{{\tt{MoCaNLO\-+RECOLA}}} 
\newcommand{\SHERPA}{{\tt{SHERPA}}} 
\newcommand{\MGAMCNLO}{{\tt{MadGraph5\_aMC@NLO}}} 
\newcommand{\MADGRAPH}{{\tt{MadGraph}}} 

\newcommand{\POWHEG}{{\tt{POWHEG}}}
\newcommand{\POWHEGBOX}{{\tt{POWHEG~BOX}}}
\newcommand{\POWHEGBOXVV}{{\tt{POWHEG~BOX~V2}}}
\newcommand{\PYTHIA}{{\tt{PYTHIA}}}
\newcommand{\PYTHIAS}{{\tt{PYTHIA6}}}
\newcommand{\PYTHIAE}{{\tt{PYTHIA8}}}
\newcommand{\HERWIG}{{\tt{HERWIG}}}
\newcommand{\HERWIGS}{{\tt{HERWIG7}}}

\newcommand{\ATLL}{{\em{ATLAS-loose}}}
\newcommand{\CMST}{{\em{CMS-tight}}}

\newcommand{\NLOPS}{{{NLO+PS}}}

\newcommand{\beq}{\begin{equation}}
\newcommand{\eeq}{\end{equation}}

\newcommand{\ed}{\end{document}}

\begin{document}
\title{ Parton-shower effects in electroweak $WZjj$ production at the
  next-to-leading order of QCD}
\author{Barbara J\"ager\thanksref{tubingen} \and Alexander Karlberg\thanksref{zurich} \and Johannes Scheller\thanksref{tubingen}}
\institute{Institute for Theoretical Physics, University of T\"ubingen,
Auf der Morgenstelle 14, 72076 T\"ubingen,  Germany \label{tubingen} \and Physics Institute, University of Z\"urich, Winterthurerstrasse 190, 8057 Z\"urich, Switzerland\label{zurich}}
\date{\today}
\journalname{EPJC}
\maketitle

\abstract{
We present an implementation of $\wzjj$ \linebreak production via vector-boson
fusion in the \POWHEGBOX{}, a public
tool for the matching of next-to-leading order QCD calculations with
multi-purpose parton-shower \linebreak generators.
We provide phenomenological results for el\-ectroweak $\wzjj$
production with fully leptonic decays at the LHC in realistic setups
and discuss theoretical uncertainties associated with the
simulation. We find that beyond the leading-order approximation the
dependence on the unphysical factorization and renormalization scales
is mild. The two tagging jets are furthermore very stable against
parton-shower effects. However, considerable
sensitivities to the shower Monte-Carlo program used are observed for
central-jet veto observables.
}

%
\section{Introduction}
%
Vector boson scattering (VBS) processes provide particularly promising
means for probing the gauge structure of the Standard Model's
electroweak sector. Processes of the type $VV\to VV$ (with $V$
denoting a $W^\pm$ or $Z$ boson) involve pure weak gauge-boson
scattering contributions as well as contributions featuring Higgs
exchange contributions. In the framework of the Standard Model (SM),
well-behaved cross sections respecting unitarity conservation up to
the highest energy scales require a subtle interplay of triple and
quartic couplings of the gauge and Higgs bosons.  Deviations in
measurements from theoretical predictions of gauge-boson scattering
processes could thus point to anomalous effects in triple and quartic
gauge boson couplings, or other types of new physics in the
electroweak sector.

At hadron colliders such as the CERN Large Hadron Collider (LHC),
on-shell gauge boson scattering processes of the type $VV\to VV$ are
not directly accessible. Instead, VBS processes are most conveniently
studied by means of the purely electroweak (EW) production processes
$pp\to VV+2~\mr{jets}$, where two protons scatter by the exchange of
weak gauge bosons that in general exhibit non-vanishing virtuality. In
addition to the two final-state gauge bosons the scattering protons
produce two so-called tagging jets that are typically located in the
forward regions of the detector.  Experimentally, such processes can
be separated rather well from strong production processes with the
same final state, because of the characteristic distribution of the
tagging jets that tend to be produced at large invariant mass via the
EW production mode, while they are mostly produced close in rapidity
with small invariant mass in QCD-dominated background processes. The
EW production mode also distinguishes itself from the QCD induced one
by generating significantly fewer jets in the detector volume between
the two tag jets~\cite{Rainwater:1999sd}. This characteristic can be used to veto such
activity and hence further reduce the QCD background.
To best exploit the characteristic VBS signature, one considers fully
leptonic decays of the gauge bosons, resulting in final states with
four leptons that are typically located between the two tagging
jets. At order $\mc{O}(\alpha^6)$ such final states cannot only stem
from genuine weak-boson scattering contributions involving a
$V^{(\star)}V^{(\star)}\to VV\to 4~\text{leptons}$ topology, but also
from other electroweak contributions not involving resonant $V$
bosons. Only the sum of all diagrams giving rise to a specific final
state with four leptons and two jets in a gauge invariant manner is
meaningful to consider.

In this work, we focus on the EW production of a final state with one neutrino, three charged leptons, and two jets, $\nu_{\ell'}
{\ell'}^+\ell^-\ell^+jj$ (with $\ell,\ell'=\mu, e$), that is dominated
by $W^+Z$ VBS contributions.
First experimental results on this production mode were reported for
an LHC center-of-mass energy of $\sqrt{s}=8$~TeV using a data sample
with an integrated luminosity of 20.3~fb$^{-1}$ by the ATLAS
collaboration in Ref.~\cite{Aad:2016ett}.
Very recently, the ATLAS and CMS collaborations have presented first
measurements of this production mode at $\sqrt{s}=13$~TeV, using data
samples collected in 2015 and 2016 corresponding to integrated
luminosities of 36.1~fb$^{-1}$ and 35.9~fb$^{-1}$,
respectively~\cite{atlas-vbs-wz13,cms-vbs-wz13}. Based on
leading-order (LO) simulations, ATLAS reports measured cross sections
slightly above the theoretical prediction \cite{atlas-vbs-wz13},
whereas in a slightly different setup CMS finds agreement of their
measurements with the SM prediction \cite{cms-vbs-wz13}.

Although current measurements are limited by sta\-tistics, the situation
will change in the future when more data become available. In
particular it will be of great importance that the available
theoretical predictions can match the expected experimental
precision~\cite{CMS:2016rcn}. It is therefore necessary to have a good
understanding of the accuracy of the particular theoretical prediction
when comparing to experimental data.

Two studies investigating the accuracy of various VBS predictions have
recently been published. The first study, a detailed analysis of the
various contributions to the $\wpmzjj$ production mode at LO, was
presented in Ref.~\cite{Bendavid:2018nar} using the Monte-Carlo
programs \MOCANLORECOLA{}~\cite{Actis:2016mpe,Actis:2012qn},
\SHERPA{}~\cite{Gleisberg:2008ta,Gleisberg:2003xi},
\MGAMCNLO{}~\cite{Alwall:2014hca}, and \VBFNLO{}~\cite{vbfnlo}. The
fixed-order parton-level comparison revealed good agreement between
the various theoretical predictions, even though only VBS-induced
contributions where considered in \VBFNLO{} while all amplitudes
contributing at order $\mc{O}(\alpha^6)$ have been retained in the
other generators. However, when the LO generators were combined with
parton-shower tools, significant scale uncertainties and larger
discrepancies between the various predictions were observed,
emphasizing the need for higher order QCD corrections at
matrix-element level. In particular observables related to the third
jet, which is not present at LO and is therefore generated purely by
the parton shower, showed very large discrepancies. Such discrepancies
are expected to decrease when predictions at next-to-leading order (NLO) of QCD are used instead.

A similar study of the related VBS process of \linebreak $W^\pm W^\pm jj$
production was presented in~Ref.~\cite{Ballestrero:2018anz}. This
process has received significantly more theoretical attention than the
other VBS processes due to its simpler structure. As such this study
did not only present results at LO, but discussed also the matching of
NLO-QCD calculations with parton showers (PS). Although it was found
that the description of all observables improves when using NLO+PS
tools compared to LO+PS it became clear that remaining uncertainties related to the parton-shower generator can be significant and should be accounted for systematically.

While NLO-QCD corrections to VBS-induced \linebreak
$\nu_{\ell'} {\ell'}^+\ell^-\ell^+jj$ production at the LHC have been
known for quite some time \cite{Bozzi:2007ur} and available in the
form of a flexible parton-level Monte-Carlo program in the \VBFNLO{}
package \cite{vbfnlo}, the matching of an NLO-QCD calculation with
parton-shower programs has not been presented so far for this channel.
Work in that direction has already been performed for the related
$\wpwpjj$, $\wpwmjj$, and $\zzjj$ VBS production modes
\cite{Jager:2011ms,Jager:2013mu,Jager:2013iza,Rauch:2016upa}. For
the $\wzjj$ mode, however, no public implementation was available up
to now. Here we close this gap by implementing the calculation of
\cite{Bozzi:2007ur} in the framework of the
\POWHEGBOX{}~\cite{powhegbox}, a tool for the matching of dedicated
NLO-QCD calculations with public parton-shower generators such as
\PYTHIA{}~\cite{Sjostrand:2006za,Sjostrand:2014zea} or
\HERWIG~\cite{Corcella:2000bw,Bellm:2015jjp}, making use of the
\POWHEG{} method~\cite{powheg}. Doing so in a public framework, we
provide the tools for performing analyses of EW $\wzjj$ production at
NLO+PS accuracy and thus avoiding theoretical uncertainties due to an
inaccurate treatment of the hard scattering process as inherent to a
mere LO simulation.

Technical details of our implementation are briefly summarized in
Sec.~\ref{sec:technical}. In Sec.~\ref{sec:pheno} we present results
of a detailed numerical analysis with particular emphasis on the
impact different parton shower generators have on phenomenological
predictions. Our conclusions are provided in Sec.~\ref{sec:concl}.

%
\section{Details of the implementation}
\label{sec:technical}
%
In order to implement the electroweak $\wzjj$ production process in
the framework of the \POWHEGBOX{} we closely followed the strategy
applied to the related VBS processes $\wpwpjj$~\cite{Jager:2011ms},
$\zjj$~\cite{Jager:2012xk}, $\wpwmjj$~\cite{Jager:2013mu}, and
$\zzjj$~\cite{Jager:2013iza}. We extracted the relevant matrix
elements for electroweak $\wzjj$ production including the fully
leptonic decays of the $W$ and $Z$ bosons at LO (i.e.\ at order
$\mc{O}(\alpha^6)$) and at NLO-QCD from \VBFNLO{} and adapted them to
comply with the format required by the \POWHEGBOXVV{}. We retained all
approximations inherent to the calculation of \cite{Bozzi:2007ur} that
forms the basis of the \VBFNLO~implementation.
%

\begin{figure*}[t!]
\centering
\includegraphics[width=0.49\textwidth]{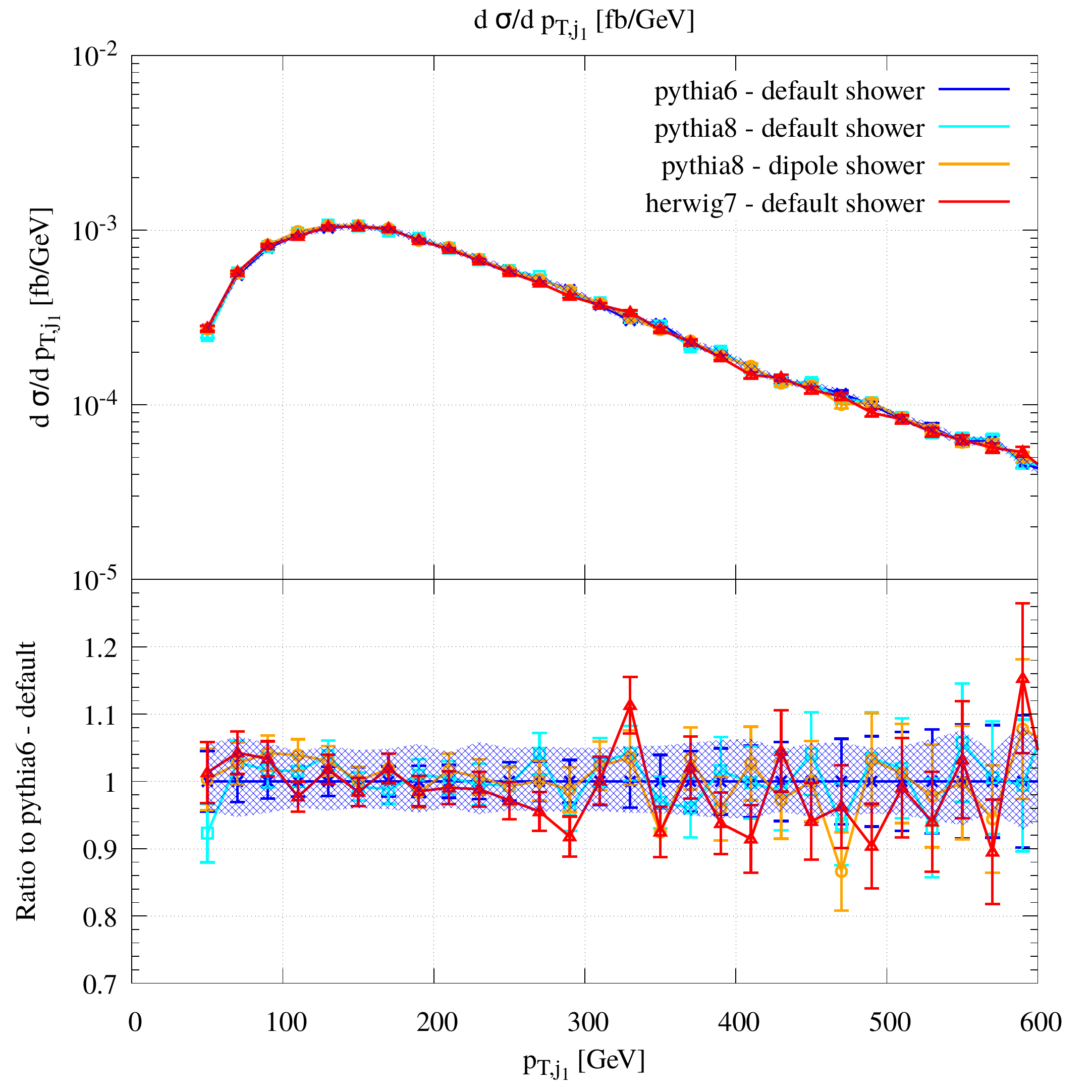}
\includegraphics[width=0.49\textwidth]{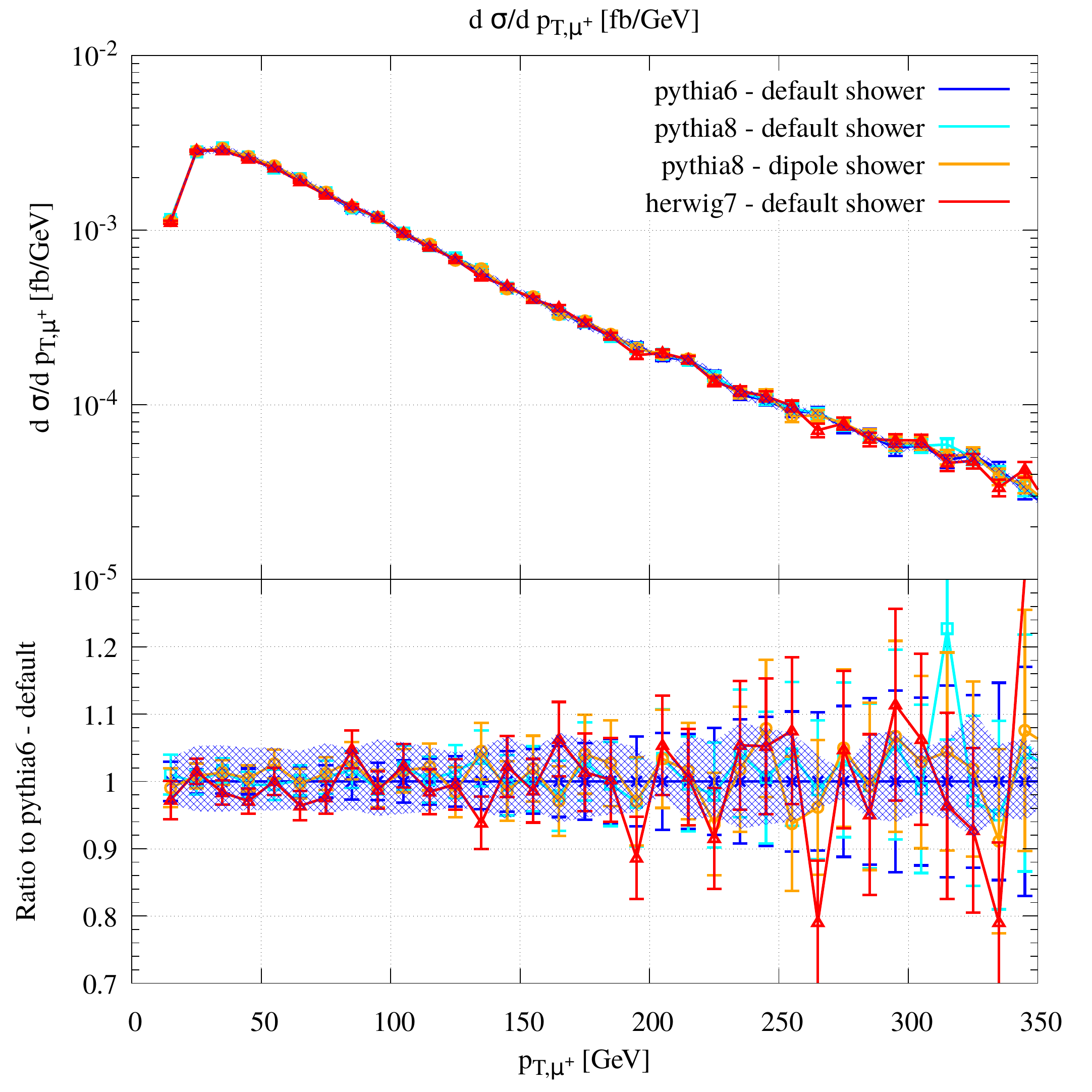}
\caption{
Transverse-momentum distributions of the hardest tagging jet (left) and of the positively charged muon (right) in the \CMST{} scenario. In the lower panels the ratio of the respective distribution to the  \PYTHIAS{} reference result is shown. In each case the blue bands indicate the scale uncertainty of the \PYTHIAS{} simulations, statistical uncertainties are denoted by error bars. 
\label{fig:ptj1-ptmu1-cms}
}
\end{figure*}

In particular, at LO we considered (anti-)quark scattering
contributions giving rise to a final state with two (anti-)quarks, a
neutrino, and three leptons, resulting in a $\emmvjj$ signature. While
for the matrix elements squared $t$-channel topologies are taken into
account as well as $u$-channel topologies, interference contributions
of $t$-channel with $u$-channel diagrams are neglected. Contributions
from $s$-channel-induced production modes are considered as part of a
three-gauge-boson production process and entirely disregarded. The
neglected contributions are small in the phase-space region where VBS
processes are typically investigated experimentally, with
widely-separated jets in the forward- and backward regions of the
detector. For instance, at LO they result in only a 0.6\% deviation
from the full calculation in the realistic setup explicitly
investigated in Ref.~\cite{Bendavid:2018nar}.  Throughout, the
Cabibbo-Kobayashi-Maskawa matrix is assumed to be diagonal, and
contributions with external top or bottom quarks are disregarded.  In
the following we will refer to electroweak $\emmvjj$ production in
proton-proton collisions with\-in the afore-mentioned approximations as
{\em VBS-induced $\wzjj$ production}.


Similar to the previously considered $\zjj$,  $\wpwmjj$, and $\zzjj$ VBS processes,
the electroweak $\wzjj$ production cross section in the fully
inclusive case exhibits various types of singularities at Born
level. While such singularities can easily be avoided by the
application of phase-space cuts in a fixed-order calculation, the
inclusive sampling of the VBS phase space required by the \POWHEGBOX{}
calls for a more subtle treatment. We achieve that goal in a two-step
procedure: First, we introduce a technical cut to remove contributions
with singularities due to the exchange of a $t$-channel photon of low
virtuality, $Q_\mr{min}^2\leq 4$~GeV$^2$. Such contributions are
entirely negligible after analysis cuts of $p_T^\mr{jet} \geq 40$~GeV
are imposed on the two tagging jets, and can therefore be disregarded
already at generation level.  Second, an improvement in the efficiency
of the phase-space integration can be achieved by applying a so-called
Born-suppression factor $F$ that dampens the integrand in singular
regions of phase space. We employ a Born-suppression factor of the
form \beq F(\Phi_n) =
\left(\frac{p_{T,1}^2}{p_{T,1}^2+\Lambda^2}\right)^2
\left(\frac{p_{T,2}^2}{p_{T,2}^2+\Lambda^2}\right)^2\,, \eeq with the
$p_{T,i}$ $(i=1,2)$ denoting the transverse momenta of the final-state
(anti-)quarks of an underlying Born configuration $\Phi_n$, and
$\Lambda$ a technical damping parameter set to 10~GeV.

In addition to singularities due to the exchange of low-virtuality
photons in the $t$-channel, in $\emmvjj$ production processes
divergences can occur when the $\mu^-\mu^+$ pair stems from the decay
of a quasi on-shell photon. In order to maintain gauge-invariance,
these diagrams have to be retained and cannot be discarded in the
treatment of the $\emmvjj$ process. Since in our
numerical analyses we will not consider QED shower effects that may
modify the kinematics of the final-state leptons, we remove such
configurations already at generation level by an explicit cut on the
invariant mass of the same-type lepton pair, retaining only events
with \beq m_{\mu^-\mu^+}>0.5~\text{GeV}\,.  \eeq

To validate our implementation, we performed several checks. We
compared results for the tree-level matrix elements squared used in
the \POWHEGBOX{} with equivalent expressions obtained by
\MADGRAPH{}~\cite{Alwall:2007st} for individual phase-space points and
found full agreement at double-precision accuracy. Additionally, we
compared parton-level results for cross sections and differential
distributions at LO and at NLO-QCD with results obtained by \VBFNLO{},
again finding full agreement within the numerical accuracy of the two
calculations.

The computer code we developed will be made available in the public
version of the \POWHEGBOXVV{}, accessible at the webpage {\tt
  http://powhegbox.mib.infn.it/}.
%
%
\begin{figure*}[thp!]
\center
\includegraphics[width=0.49\textwidth]{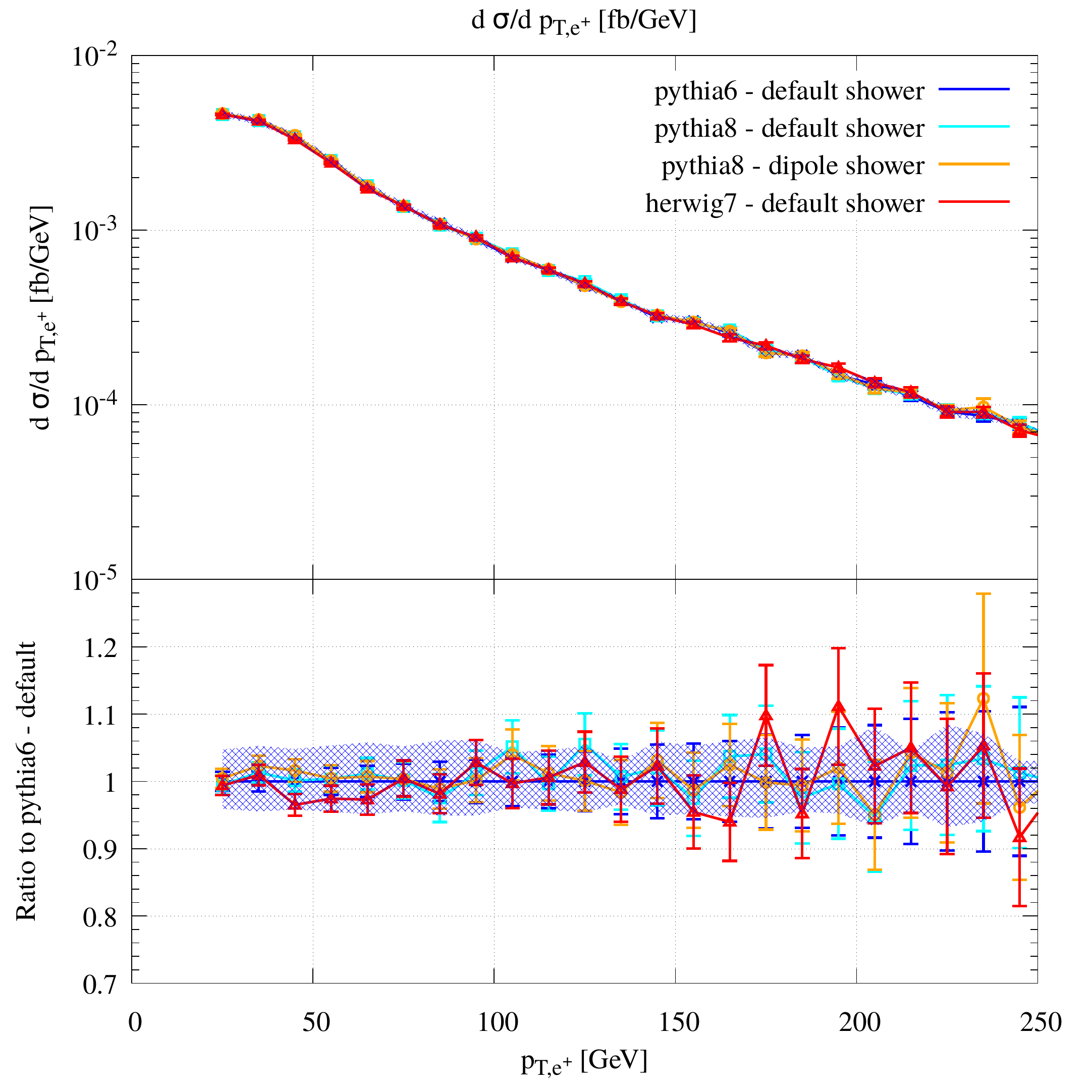}
\includegraphics[width=0.49\textwidth]{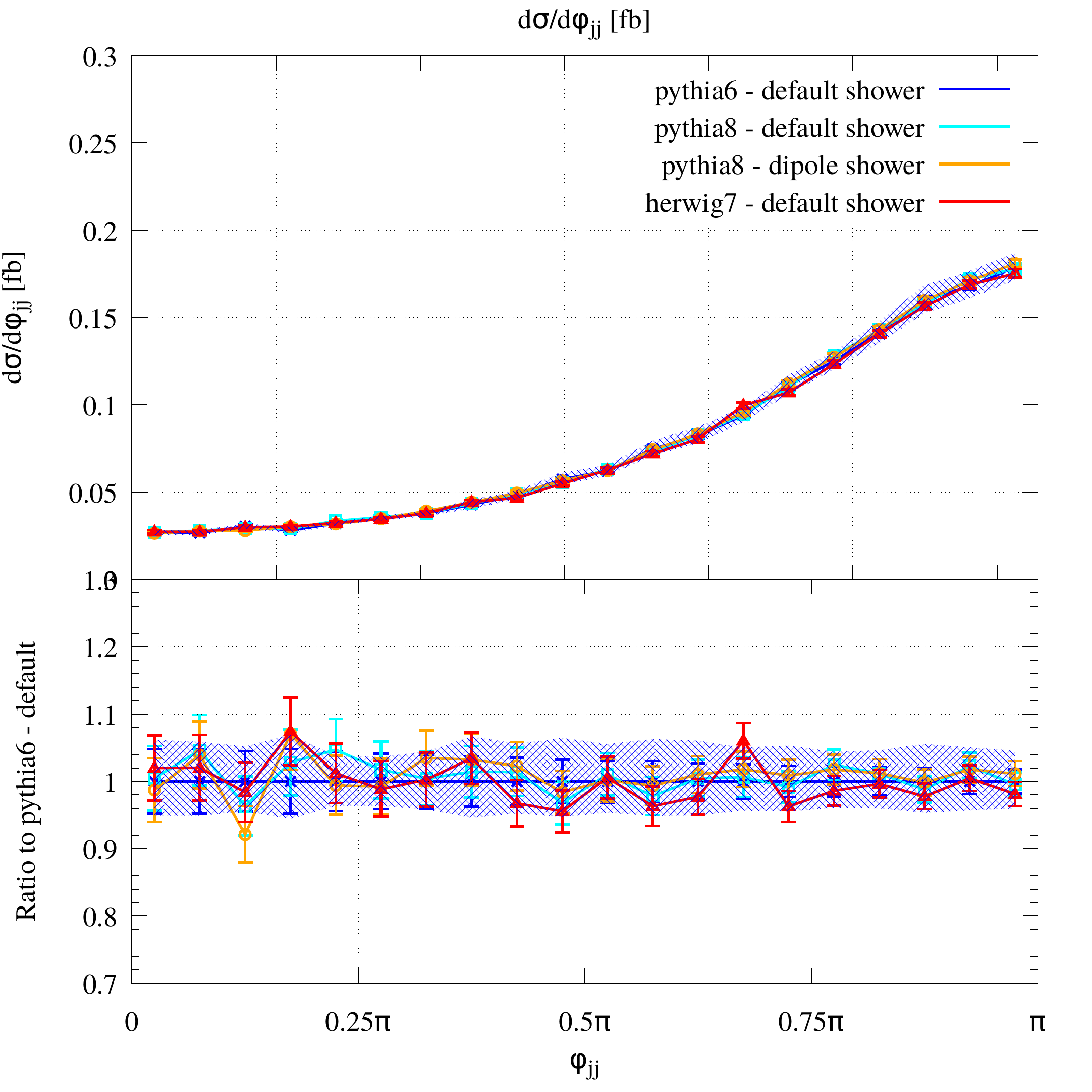}
\caption{
Transverse-momentum distribution of the hardest positron (left) and azimuthal-angle separation of the two tagging jets (right) in the \CMST{} scenario. In the lower panels the ratio of the respective distribution to the  \PYTHIAS{} reference result is shown. In each case the blue bands indicate the scale uncertainty of the \PYTHIAS{} simulations, statistical uncertainties are denoted by error bars.  
\label{fig:pte1-phijj-cms}
}
\end{figure*}
\section{Phenomenological results}
\label{sec:pheno}
In order to demonstrate the capabilities of the developed code we present phenomenological results for a few selected setups inspired by realistic experimental studies. 

In each case, we consider proton-proton collisions at the LHC with a
center-of-mass energy of $\sqrt{s}=13$~TeV. We use the PDF4LHC15
parton distribution function (PDF) set at
NLO~\cite{Butterworth:2015oua} as provided by the LHAPDF
library~\cite{Buckley:2014ana} (LHAPDF ID $90000$). For the definition
of jets we employ the anti-$k_T$
algorithm~\cite{Cacciari:2005hq,Cacciari:2008gp} with $R=0.4$, as
implemented in the {\tt FASTJET 3.3.0} package~\cite{Cacciari:2011ma}. The mass and width of the
Higgs boson are set to $m_H=125.18$~GeV and $\Gamma_H=0.00407$~GeV,
respectively. As EW input parameters we chose the Fermi constant,
$G_\mu=1.16638\times 10^{-5}$~GeV$^{-2}$, and the masses of the $W$
and $Z$ bosons, $m_W=80.379$~GeV and $m_Z=91.188$~GeV,
respectively. The widths of the massive gauge bosons are set to
$\Gamma_W=2.085$~GeV and $\Gamma_Z=2.4952$~GeV.

We provide results for a setup inspired by the ATLAS analysis of Ref.~\cite{atlas-vbs-wz13}, dubbed \ATLL, and for a scenario with more stringent cuts, referred to as \CMST, following the CMS analysis of Ref.~\cite{cms-vbs-wz13}. 
In order not to overload the discussion in the main body of this article, in this section we focus on results in the \CMST{} scenario. For reference, we show the same distributions obtained in the \ATLL{} scenario in Appendix~\ref{appendix}.

%
For the \ATLL{} scenario, we require the presence of at least two 
jets with transverse momenta and rapidities of 
\beq
p_{T,j}>40~\mr{GeV}\,,\quad
|y_j|<4.5\,, 
\eeq
respectively. 
The two hardest jets satisfying these criteria are called ``tagging jets'', and are required to  exhibit an invariant mass 
\beq
m_{j_1j_2}> 150~\mr{GeV}\,.
\eeq
In addition to the tagging jets each signal event is supposed to contain an $e^+$, a $\mu^+$, and a $\mu^-$ with 
\beq
|y_\ell|<2.5\,,\quad
\Delta R_{\ell\ell}>0.2\,,\quad
\Delta R_{j\ell}>0.2\,,
\eeq
where $\ell$ generically denotes a charged lepton and $\Delta R_{ij}$ the separation of two particles $i,j$ in the rapidity--azi\-muthal angle plane. The muons are required to be close in invariant mass to the $Z$-boson resonance with 
\beq
|m_Z-m_{\mu^+\mu^-}|<10~\mr{GeV}\,,\quad 
p_{T,\mu}>15~\mr{GeV}\,,\quad
\eeq
while a harder transverse-momentum cut is imposed on the electron, 
\beq
p_{T,e}>20~\mr{GeV}\,.
\eeq

The \CMST{} scenario comprises the following stringent selection cuts on the jets with 
\beq
p_{T,j}>50~\mr{GeV}\,,\quad
|y_j|<4.7\,. \quad
\eeq
The two hardest jets satisfying these criteria are called ``tagging jets'', and are furthermore required to fulfill the additional requirements 
\beq
m_{j_1j_2}> 500~\mr{GeV}\,,\quad
|\Delta y_{j_1j_2}|=|y_{j_1}-y_{j_2}|>2.5\,,
\eeq
supplemented by the separation cuts 
\beq
\Delta R_{j\ell}>0.4\,,\quad
\Delta R_{\ell\ell}>0.2\,,
\eeq
and  
cuts specifically distinguishing between the hardest and second-hardest muon of each event,
\beq
p_{T,\mu_1}>25~\mr{GeV}\,,\quad
p_{T,\mu_2}>15~\mr{GeV}\,,\quad
\eeq
the electron, and the missing transverse momentum which in our calculation corresponds to the momentum of the neutrino
\beq
p_{T,e}>20~\mr{GeV}\,,\quad
p_T^\mr{miss}>30~\mr{GeV}\,,
\eeq
with pseudorapidities of 
\beq
|\eta_\mu|<2.4\,,\quad
|\eta_e|<2.4\,,\quad
\eeq
and cuts on the invariant masses of the $\mu^+\mu^-$ and the three-lepton systems, respectively, 
\beq
m_{\mu^+\mu^-}>4~\mr{GeV}\,,\quad
m_{e^+\mu^+\mu^-}>10~\mr{GeV}\,.\quad
\eeq
In addition, the location of the three-lepton system 
in pseudo-rapidity, $\eta_{3\ell}$, with respect to the two tagging jets is required to fulfill
\beq
|\eta_{3\ell}-\frac{\eta_{j_1}-\eta_{j_2}}{2}|<2.5\,.
\eeq
We note that when considering distributions of the third jet in the following, we only consider events with $p_{T,j_3}> 10~\mr{GeV}$, as softer jets cannot be reliably identified. For the discussion of the transverse-momentum distribution of the third jets, naturally no such requirement is imposed. For clarity, whenever we restrict the considered range of $p_{T,j_3}$ we will specifically indicate that in the respective figures. 

The factorization and renormalization scales, $\muf$ and $\mur$, 
for each phase-space point are identified with the geometric mean of the
transverse momenta of the two final-state partons of the underlying
Born configuration, \beq \mu_0\equiv\sqrt{p_{T,1}\cdot p_\mr{T,2}}\,.
\eeq
For studying the dependence of the various results on these scales, we
introduce the two parameters $\xif$ and $\xir$ such that
$\muf = \xif\mu_0$ and $\mur = \xir\mu_0$, and vary them
independently.

In order to explore the impact of parton-shower effects on
experimentally accessible distributions and to investigate their
uncertainties, we consider three different parton-shower Monte-Carlo
generators (SMCs): \PYTHIA~6.4.28~\cite{Sjostrand:2006za},
\PYTHIA~8.230~\cite{Sjostrand:2014zea}, and
\HERWIG~7.1.4 \cite{Bellm:2015jjp}. For \PYTHIAE{} we use the Monash
Tune ({\texttt Tune:pp = 14})~\cite{Skands:2014pea}, for \PYTHIAS{} we
use the {\texttt Perugia 2012 M8LO} tune~\cite{Skands:2010ak} and for
\HERWIGS{} we use the default tune. In particular, \PYTHIAE{} and
\HERWIGS{} offer the possibility of a local recoil dipole shower as an
alternative to their respective default global recoil showers. The
local recoil schemes have been shown to describe data for
deep-inelastic scattering~\cite{Cabouat:2017rzi} well, and have also
been found to be advantageous for the theoretical description of VBS
$W^\pm W^\pm jj$ production~\cite{Rauch:2016upa} and for $t$-channel
top-quark production~\cite{Carrazza:2018mix}.  Both of these processes
are similar to the VBS $\wzjj$ process from the point of view of the
parton~shower as they share the same color flow between the initial
and final state partons at LO. In order to explore the possible
advantages of a dipole shower in the description of VBS $\wzjj$
production we therefore compare results obtained with the default
recoil showers of the various SMCs (\PYTHIAS{}, \HERWIGS{}, and
\PYTHIAE{}) to the representative dipole shower of
\PYTHIAE~\cite{Cabouat:2017rzi}.

\begin{figure*}[tp]
	\centering
	\includegraphics[width=0.49\textwidth]{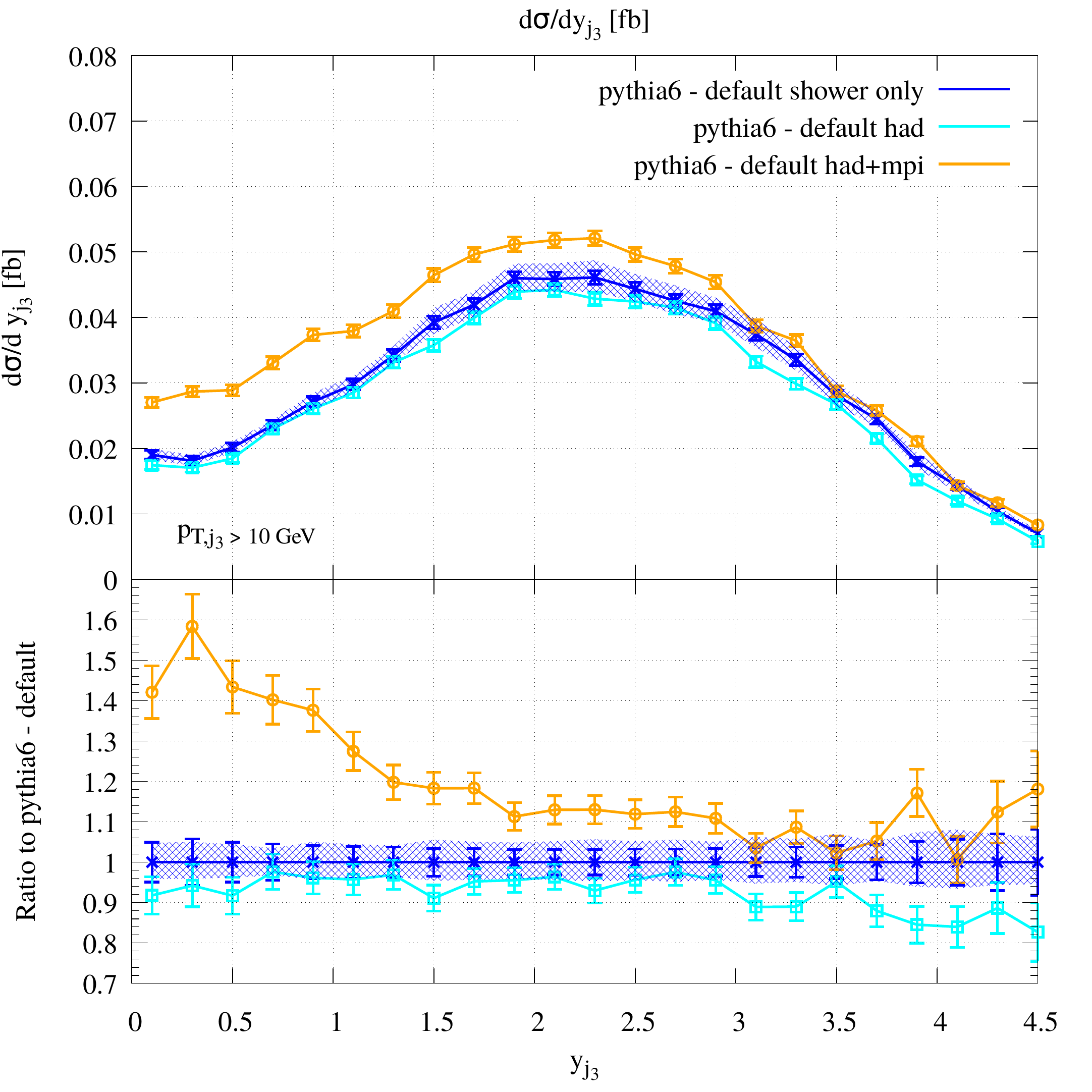}
	\includegraphics[width=0.49\textwidth]{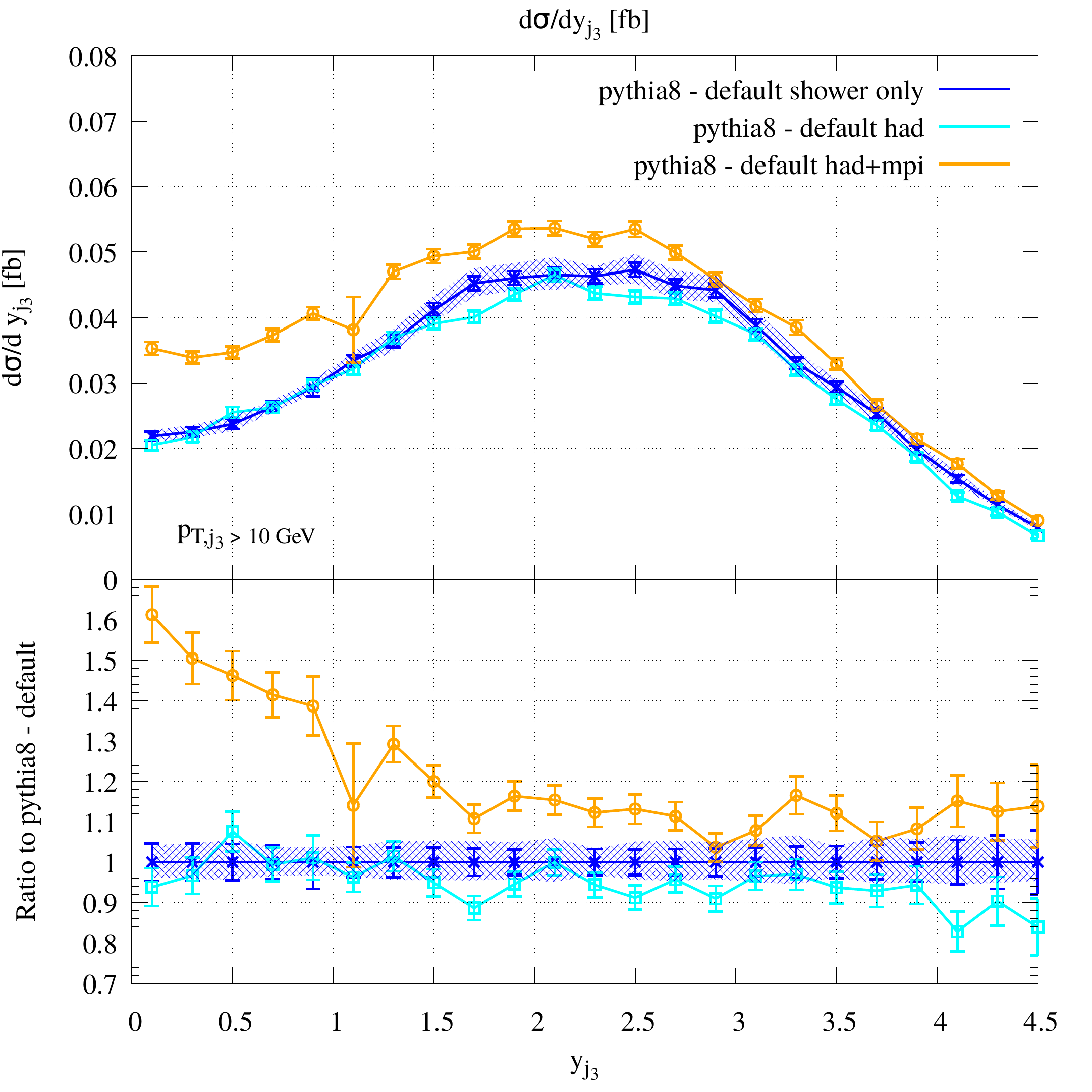}
	\includegraphics[width=0.49\textwidth]{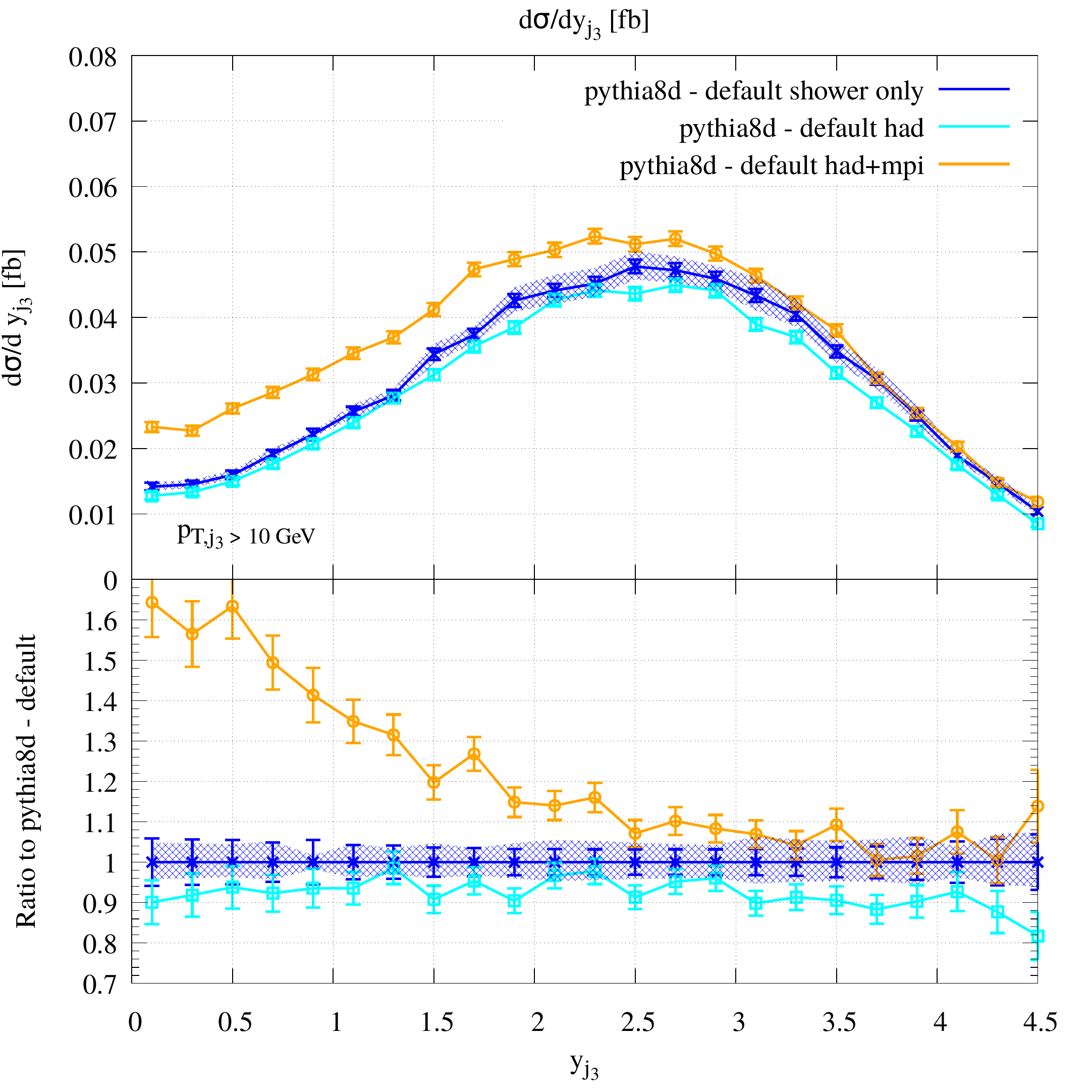}
	\includegraphics[width=0.49\textwidth]{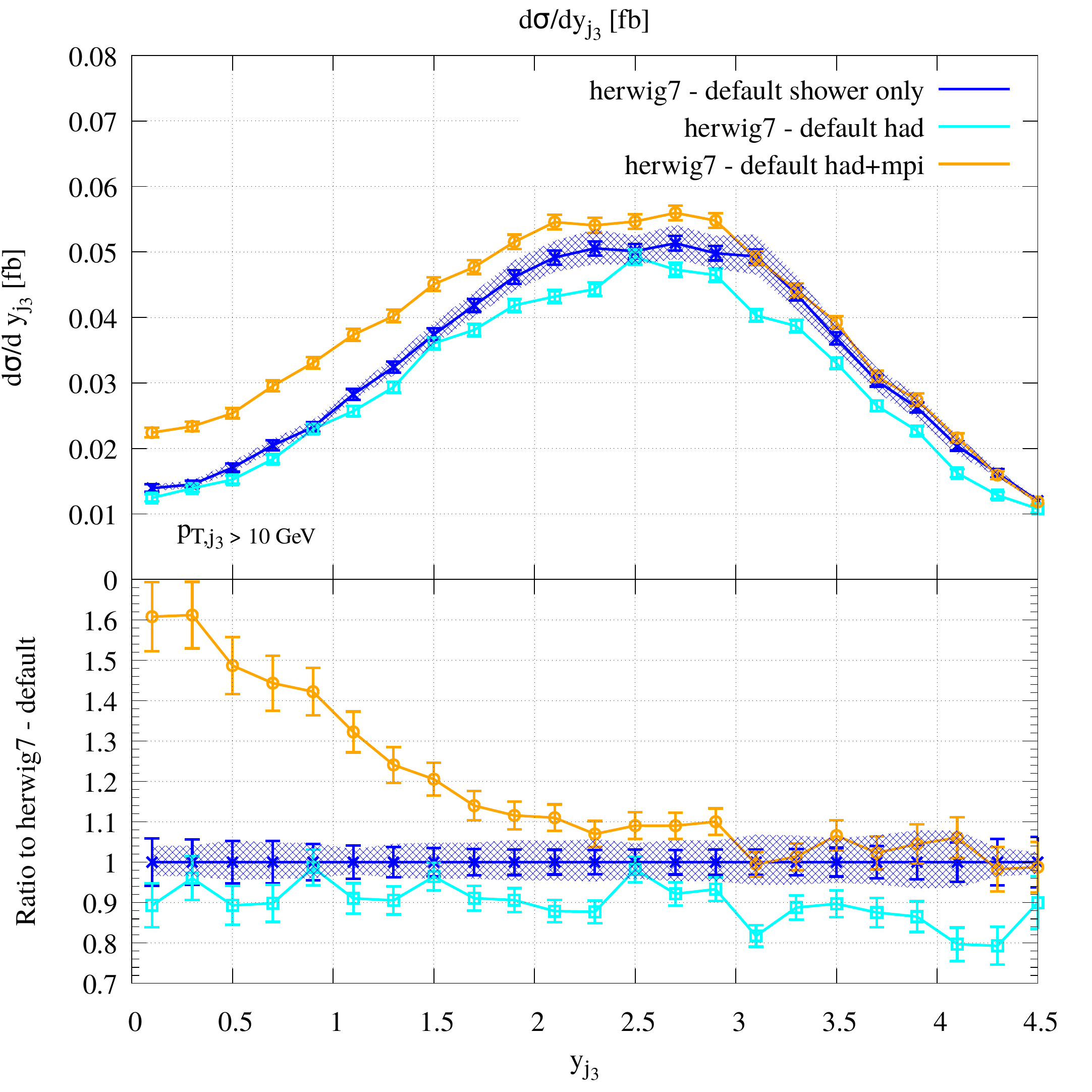}
\caption{Rapidity distribution of the third jet in the \CMST{} scenario for different SMC settings [red: parton shower only, light blue: parton shower with hadronization, dark blue: parton shower with hadronization and multi-parton  interactions]. The same observable is shown for \texttt{PYTHIA6} (a), \texttt{PYTHIA8} with recoil shower (b), \texttt{PYTHIA8} with dipole shower (c), and \texttt{HERWIG7} (d). 
 For each SMC, the lower panel shows the ratio $R_\mr{SMC}=d\sigma(\mr{SMC_{option}})/d\sigma(\mr{SMC_{default}})$ of the respective distribution. The blue bands indicate the scale uncertainty of the respective simulations without hadronization and MPI, statistical uncertainties are denoted by error bars. 
\label{fig:all-yj3-cms}}
\end{figure*}

Since the focus of our own work was on the inclusion of the
perturbative part of fixed-order calculation and parton shower, we
first show a set of predictions in a setup where perturbative effects
are not superimposed by potentially significant hadronsation effects
that are due to intrinsic details of a specific shower Monte Carlo
program. In order to quantify the impact of hadronisation effects and
multi-parton interactions we have, however, additionally investigated
distributions in a setup where these effects are fully taken into
account (c.f. figs.~\ref{fig:all-yj3-cms} and
\ref{fig:all-yj3-atlas}).

Typically, distributions of the tagging jets and of the leptons are very stable with respect to parton-shower effects. This feature is illustrated for the transverse-momentum distributions of the hardest jet and of the positively-charged muon in the \CMST{} setup in fig.~\ref{fig:ptj1-ptmu1-cms}, 
and for the transverse-momentum distribution of the hardest positron and the azimuthal-angle separation of the two tagging jets in fig.~\ref{fig:pte1-phijj-cms}. 

In these figures \NLOPS{} results are presented for all SMCs considered in this work. We show the distributions per~se for each SMC, as well as their ratio to our reference result obtained with  \PYTHIAS{}, 
\beq R = \frac{d\sigma(\mr{SMC})}{d\sigma(\PYTHIAS(\mu_0))}\,.  \eeq
In order to illustrate the scale uncertainty of our predictions, for
the \PYTHIAS{} reference results we have performed a seven-point
variation of the scale parameters $\xif, \xir$ -- that is the two
parameters $\xif$ and $\xir$ have been varied independently from 0.5
via 1 to 2 dropping configurations where $\xif$ and $\xir$ differ by a
factor larger than two. The resulting range of results is indicated by
the blue shaded band in the respective distributions.
%
\begin{figure*}[thp!]
\center
\includegraphics[width=0.49\textwidth]{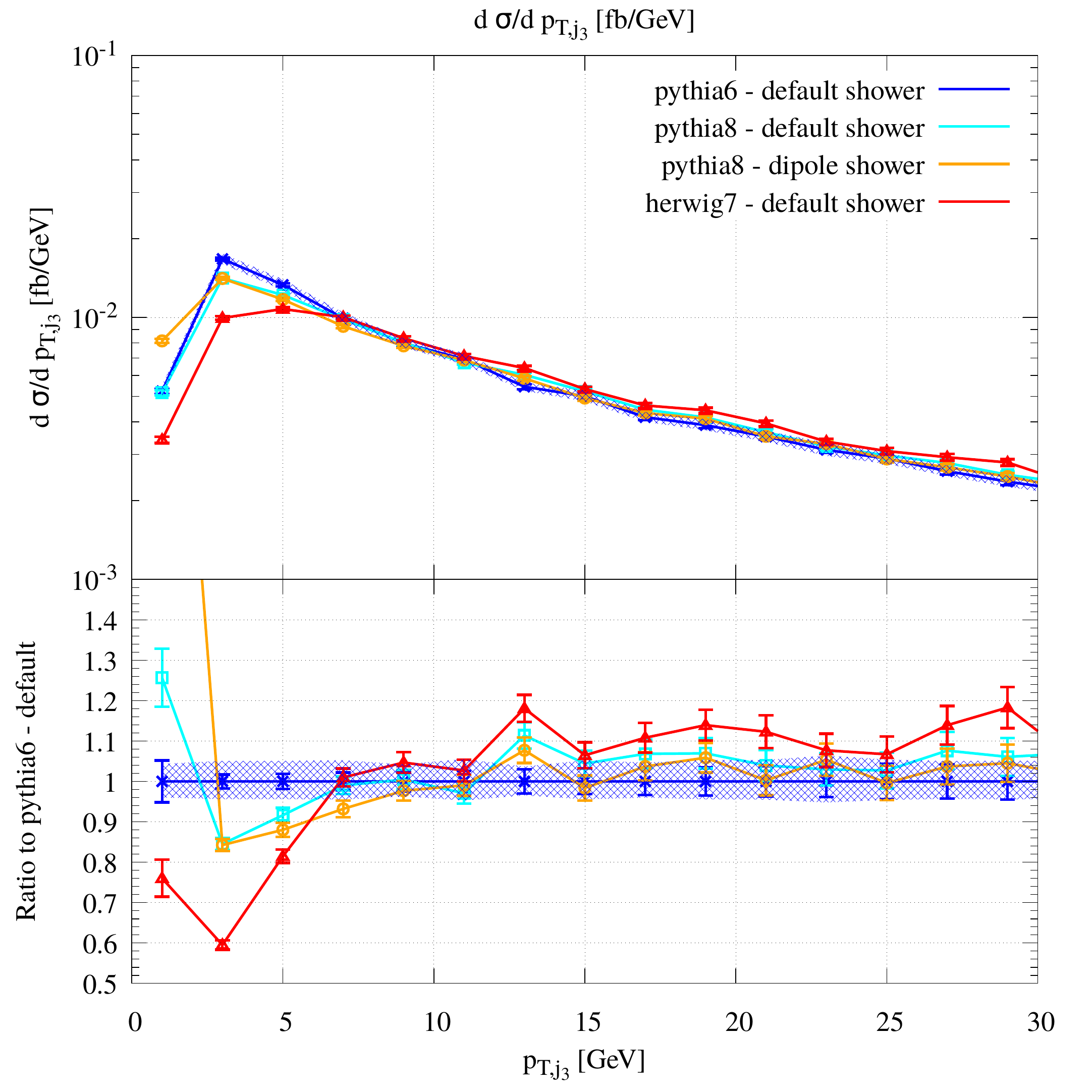}
\includegraphics[width=0.49\textwidth]{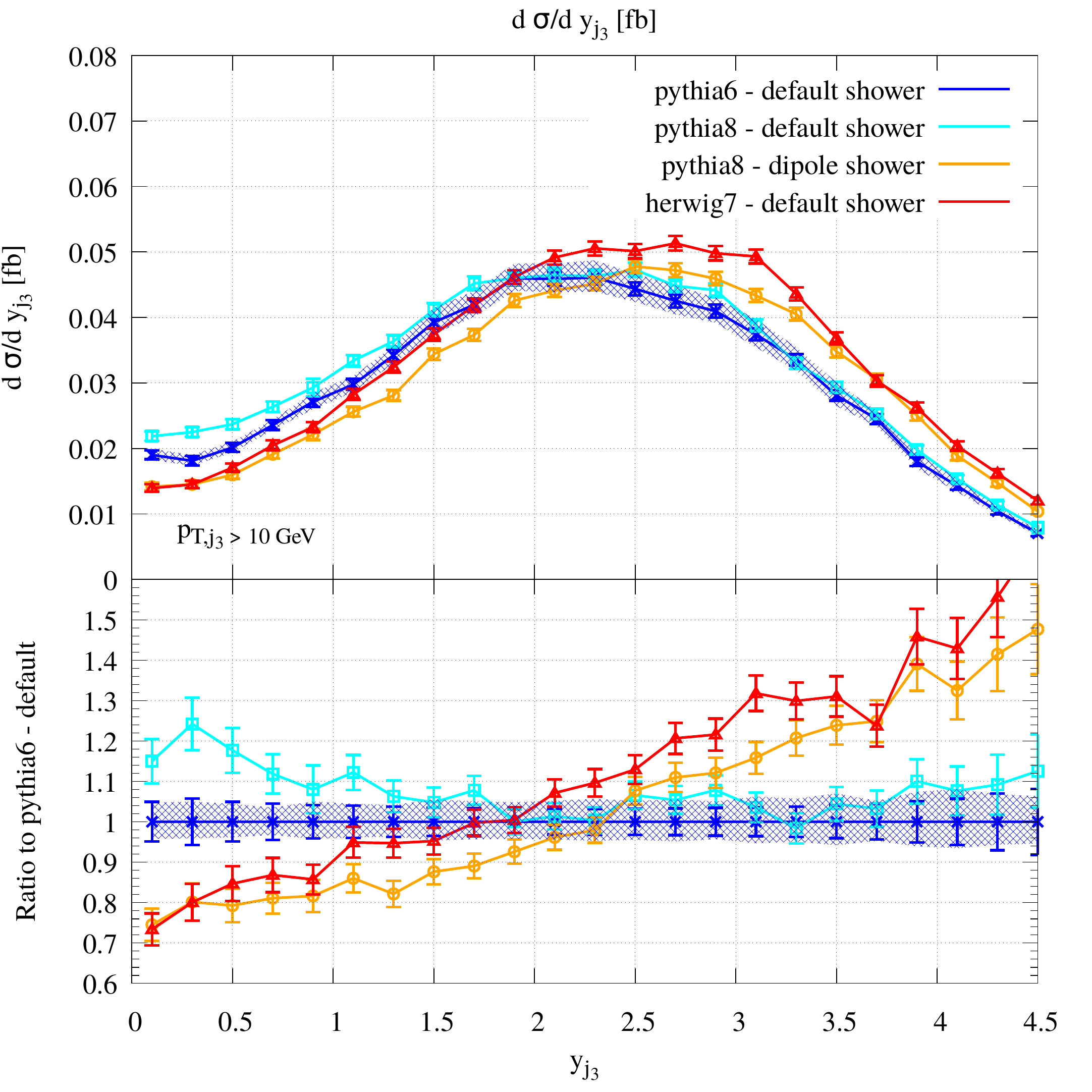}
\caption{ 
Transverse-momentum (left) and rapidity (right) distributions of the third jet in the \CMST{} scenario. In the lower panels the ratio of the respective distribution to the  \PYTHIAS{} reference result is shown. In each case the blue bands indicate the scale uncertainty of the \PYTHIAS{} simulations, statistical uncertainties are denoted by error bars. 
\label{fig:ptj3-yj3-cms}
}
\end{figure*}
%
We notice that differences due to the four parton shower algorithms
used are of the order of or even smaller than the theoretical
uncertainty due to scale variation, which amounts to about $5~\%$. As
expected, the choice of SMC does not have a significant impact on the
normalization and shape of the considered distributions.

Much larger dependencies on the used SMC are however found for distributions of the non-tagging jets.  We note that in the NLO-QCD calculation for VBS $\wzjj$ production the third jet enters only via the real-emission corrections and thus is accounted for effectively only with LO accuracy. In the \NLOPS{} calculation a third jet can also stem from the parton shower.  
Figure~\ref{fig:ptj3-yj3-cms} shows the transverse-momentum and the rapidity distributions of the third jet in the \CMST{} scenario. 

While the scale uncertainty of these distributions still amounts to
less than $10\,\%$, the differences among the various SMCs can be
significantly larger. In the transverse-momentum distribution, the
differences between predictions obtained with different versions of
\PYTHIA{} and \HERWIG{} are very large in the low-$p_T$ region, where
the fixed-order calculation is not reliable because of infrared
divergent configurations. The Sudakov factor of the {\tt NLO+POWHEG}
calculation dampens the would-be divergences resulting in a
well-behaved shape of the distribution even in the low-$p_T$
region. In the tail of the distribution the differences between the
various SMCs disappear, making predictions in the range above
$p_{T,j_3}\gtrsim 20$~GeV, where jets can be identified experimentally,
more reliable.

The differences between the individual SMCs very much affect also the rapidity distribution of the third jet. In particular, \PYTHIA{} simulations with the default recoil shower tend to fill the central-rapidity region much more than \HERWIGS{} with its default recoil shower, while \PYTHIAE{} with the dipole shower resembles the \HERWIGS{} result. 
These SMC dependencies should be kept well in mind when using so-called {\em central-jet vetoing} (CJV)  techniques.

An enhancement of the VBS signal with respect to QCD-dominated background processes can be achieved by exploiting the typical distribution of sub-leading jets depending on the nature of the production process. QCD-induced processes typically exhibit large jet activity in the central region of rapidity. In VBS processes, on the other hand, extra jets tend to be produced close to the tagging jets in the forward and backward regions of the detector. Thus, vetoing events with hard central jets (emerging in addition to the two tagging jets) helps to reduce the impact of unwanted background processes on the VBS signal.

\begin{figure*}[t]
	\centering
	\includegraphics[width=0.49\textwidth]{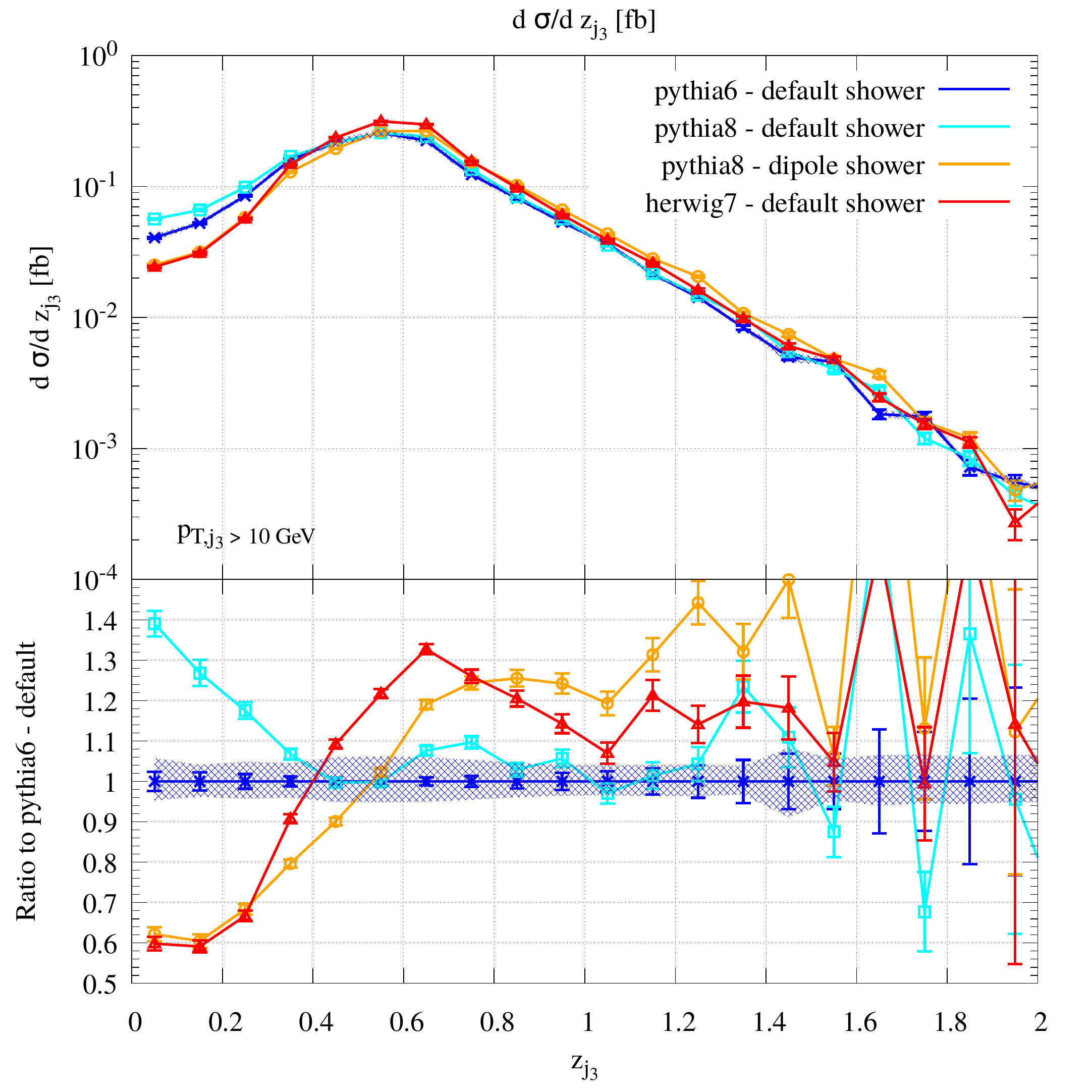}
	\includegraphics[width=0.49\textwidth]{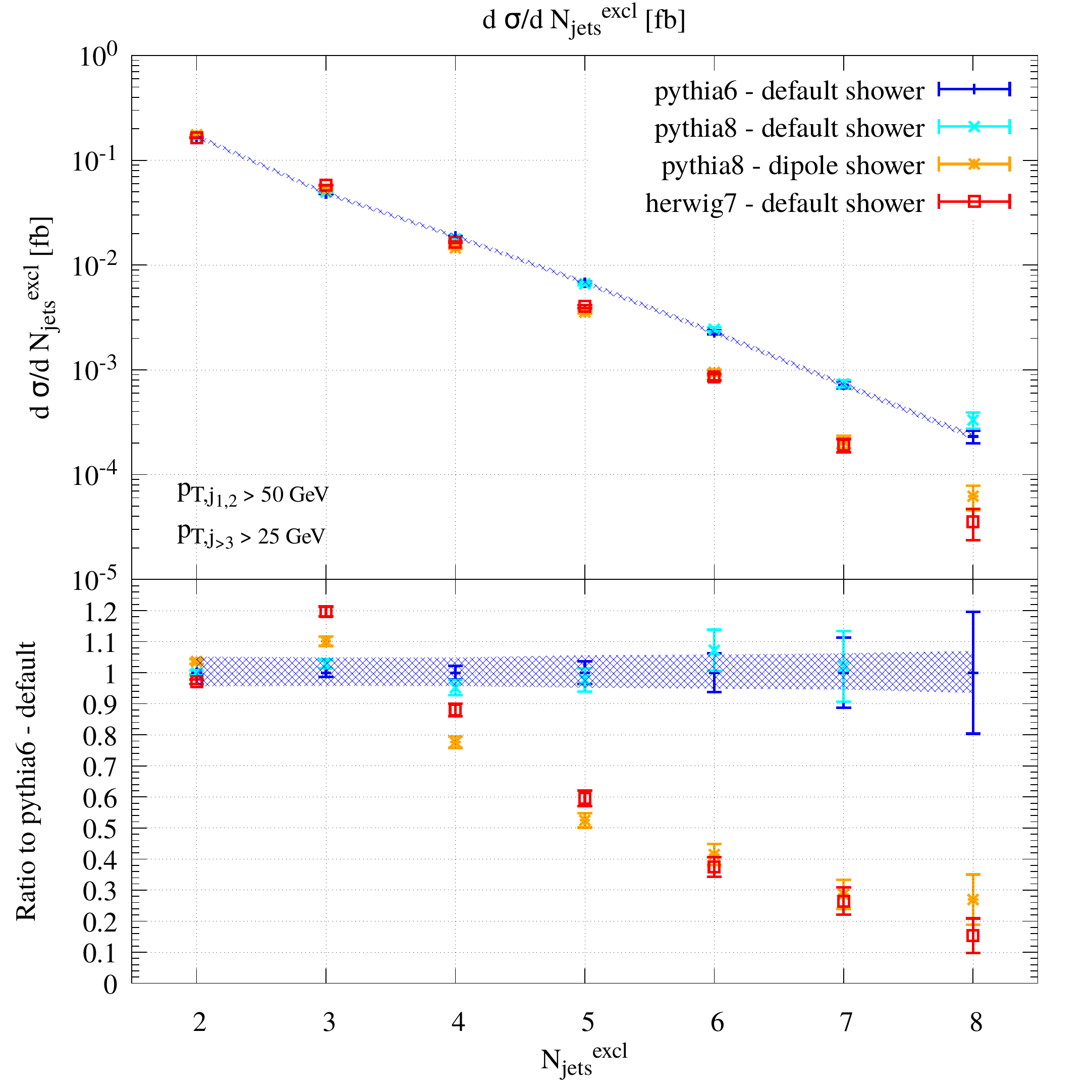}
	\caption{Differential distribution of the Zeppenfeld variable defined in eq.~\eqref{Zeppenfeld} (left) and the number of exclusive jets (right) in the \CMST{} scenario. In the left plot we require $p_{T,j_3} > 10$ GeV and in the right plot we require all jets in addition to the two tag jets to have $p_T > 25$ GeV. In the lower panel the ratio of the respective distribution to the  \PYTHIAS{} reference result is shown. In each case the blue bands indicate the scale uncertainty of the \PYTHIAS{} simulations, statistical uncertainties are denoted by error bars.
          \label{fig:PSZj3}}
\end{figure*}

The relative position of the third jet with respect to the center of the tagging-jet system is accounted for by the so-called {\em Zeppenfeld variable}, 
\beq
\label{Zeppenfeld}
z_{j_3}=\frac{y_{j_3}-\frac{y_{j_1}+y_{j_2}}{2}}{\left|\Delta y_{j_1j_2}\right|}\,,
\eeq
which approaches zero when the third jet is located just in the center of the two tagging jets, and one half, if it is close to one of the tagging jets, while values of $z_{j_3}$ larger than~0.5 indicate that the third jet is located outside of the rapidity interval spanned by the two tagging jets.
Similarly to $y_{j_3}$, this quantity helps to distinguish VBS-induced signal processes from QCD-dominated backgrounds, as much more jet activity in the rapidity region between the two tagging jets is expected for the QCD-induced production of a specific final state. 
Our results for the distribution of $z_{j_3}$ in the \CMST{} scenario are shown in fig.~\ref{fig:PSZj3}, 
%
%
%
indicating that the third jet prefers to be close to either of the tagging jets. We observe relatively large differences between the predictions obtained with the various parton showers we considered, which reflect the differences we already observed for the rapidity of the third jet entering the definition of the $z_{j_3}$ variable. This relatively large SMC uncertainty has to be kept in mind when the Zeppenfeld variable is considered as a discriminating variable between QCD- and VBS-induced processes.

In the same figure we also show the number of exclusive jets ($N_{\mathrm{jets}}^{\mathrm{excl}}$). The jets in addition to the two tag jets are required to have $p_T > 25$ GeV. We see that the various SMCs agree very well and within the scale variation band for $N_{\mathrm{jets}}^{\mathrm{excl}} = 2$, whereas for $N_{\mathrm{jets}}^{\mathrm{excl}} \ge 3$ large discrepancies are found. In particular for the jets which are dominated by the parton shower we see discrepancies of $\mathcal{O}(100\%)$.   

Up to this point, we only examined differences between predictions obtained with different parton shower algorithms, but not differences due to the settings of one and the same parton-shower program. However, it is interesting also to explore the impact of hadronization and multi-parton interactions (MPI) on predictions obtained with a specific SMC. Since our previous discussion revealed that significant differences on observables in VBS processes are to be expected mostly for distributions related to the non-tagging jets, we will restrict this comparison to those. 

Interestingly, the differences due to the settings of a specific SMC are larger than one might naively expect. This is illustrated for the rapidity distribution of the third jet in fig.~\ref{fig:all-yj3-cms}.  
%
%
%
While switching on the hadronization procedure already affects the results by an order of about  $10~\%$, allowing for multiple parton interactions enhances the cross section even more significantly. 
This effect is particularly pronounced in the central-rapidity region most relevant for central-jet vetoing techniques.

%
\section{Summary and conclusions}
\label{sec:concl}
In this article we presented an implementation of VBS-induced $\wzjj$
production in the framework of the \linebreak \POWHEGBOXVV. To
illustrate the capabilities of the developed code, we showed results
for EW $\emmvjj$ production in realistic setups, inspired by two
recently presented experimental
analyses~\cite{atlas-vbs-wz13,cms-vbs-wz13}. In particular, we
investigated the impact of different parton-shower Monte-Carlo
programs on experimentally accessible observables. We found the
tagging jets to be rather insensitive to the specific parton-shower
program used. In particular the spread in predictions using different
SMCs is usually covered by the residual scale uncertainty associated
with the NLO calculation in the framework of the {\tt POWHEG} matching
framework we considered.  Since {\tt POWHEG} generates the first emission
with its own internal Sudakov form factor, the effect of the parton
shower starts at the second emission.  We note, however, that in
principle larger differences might occur when other matching schemes
are used.

The description of subleading jets is plagued by larger uncertainties
that are often outside the scale uncertainty band. The various SMCs
differ widely in how they fill the volume between the two tag jets
with softer emissions, as can be seen by inspecting the $z_{j_3}$
distribution. We therefore strongly recommend the use of NLO+PS
simulations in analyses making use of central-jet veto techniques,
preferably comparing the outputs of more than one SMC, as calculations
based on an LO approximation for a typical VBS signature are not
capable of accurately describing the kinematics of central jets.

We also investigated the impact of hadronisation and multiple parton
interactions on the third jet observables. Although these effects are
more modest than the spread in SMCs, they can still amount to
$\sim10~\%$, a fact which should also be taken into account in
central-jet veto analyses.

The computer code we developed will be made available in the public
version of the \POWHEGBOXVV{}, accessible at the webpage {\tt
  http://powhegbox.mib.infn.it/}.

%
%
\begin{acknowledgement}
We thank Tomáš Ježo for help with interfacing the \POWHEGBOX{} and
\texttt{HERWIG7}. We are grateful for a careful reading of the
manuscript to Giulia Zanderighi.  The work of B.~J.\ and J.~S.\ has
been supported in part by the Institutional Strategy of the University
of T\"ubingen (DFG, ZUK 63), and in part by the German Federal
Ministry for Education and Research (BMBF) under grant number
05H15VTCAA.  The work of A.~K.\ was supported by the Swiss National
Science Foundation (SNF) under grant number 200020-175595.
The authors acknowledge support by the state of Baden-W\"urttemberg
through bwHPC and the German Research Foundation (DFG) through grant
no INST 39/963-1 FUGG. 
\end{acknowledgement}


%
\appendix
\section{Appendix: Results in the ATLAS-loose setup}
\label{appendix}
In this appendix we show plots in the \emph{ATLAS-loose} scenario. The
observables shown are exactly the same as the ones shown in the main
body of the paper for the \emph{CMS-tight} scenario.

For the \ATLL{} scenario, we require the presence of at least two 
jets with transverse momenta and rapidities of 
\beq
p_{T,j}>40~\mr{GeV}\,,\quad
|y_j|<4.5\,, 
\eeq
respectively. 
The two hardest jets satisfying these criteria are called ``tagging jets'', and are required to  exhibit an invariant mass 
\beq
m_{j_1j_2}> 150~\mr{GeV}\,.
\eeq
In addition to the tagging jets each signal event is supposed to contain an $e^+$, a $\mu^+$, and a $\mu^-$ with 
\beq
|y_\ell|<2.5\,,\quad
\Delta R_{\ell\ell}>0.2\,,\quad
\Delta R_{j\ell}>0.2\,,
\eeq
where $\ell$ generically denotes a charged lepton and $\Delta R_{ij}$ the separation of two particles $i,j$ in the rapidity--azi\-muthal angle plane. The muons are required to be close in invariant mass to the $Z$-boson resonance with 
\beq
|m_Z-m_{\mu^+\mu^-}|<10~\mr{GeV}\,,\quad 
p_{T,\mu}>15~\mr{GeV}\,,\quad
\eeq
while a harder transverse-momentum cut is imposed on the electron, 
\beq
p_{T,e}>20~\mr{GeV}\,.
\eeq

\begin{figure*}[tp]
\center
\includegraphics[width=0.49\textwidth]{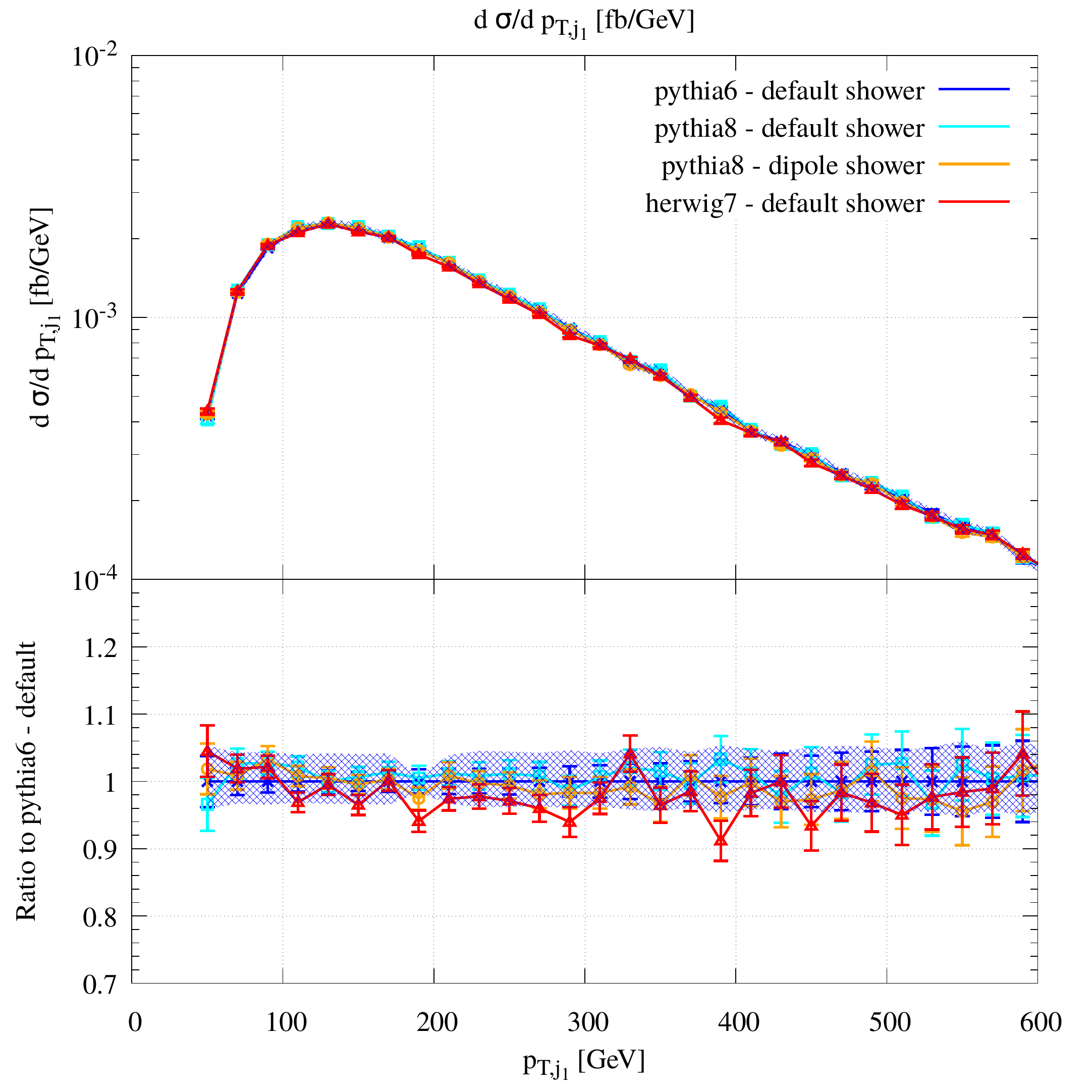}
\includegraphics[width=0.49\textwidth]{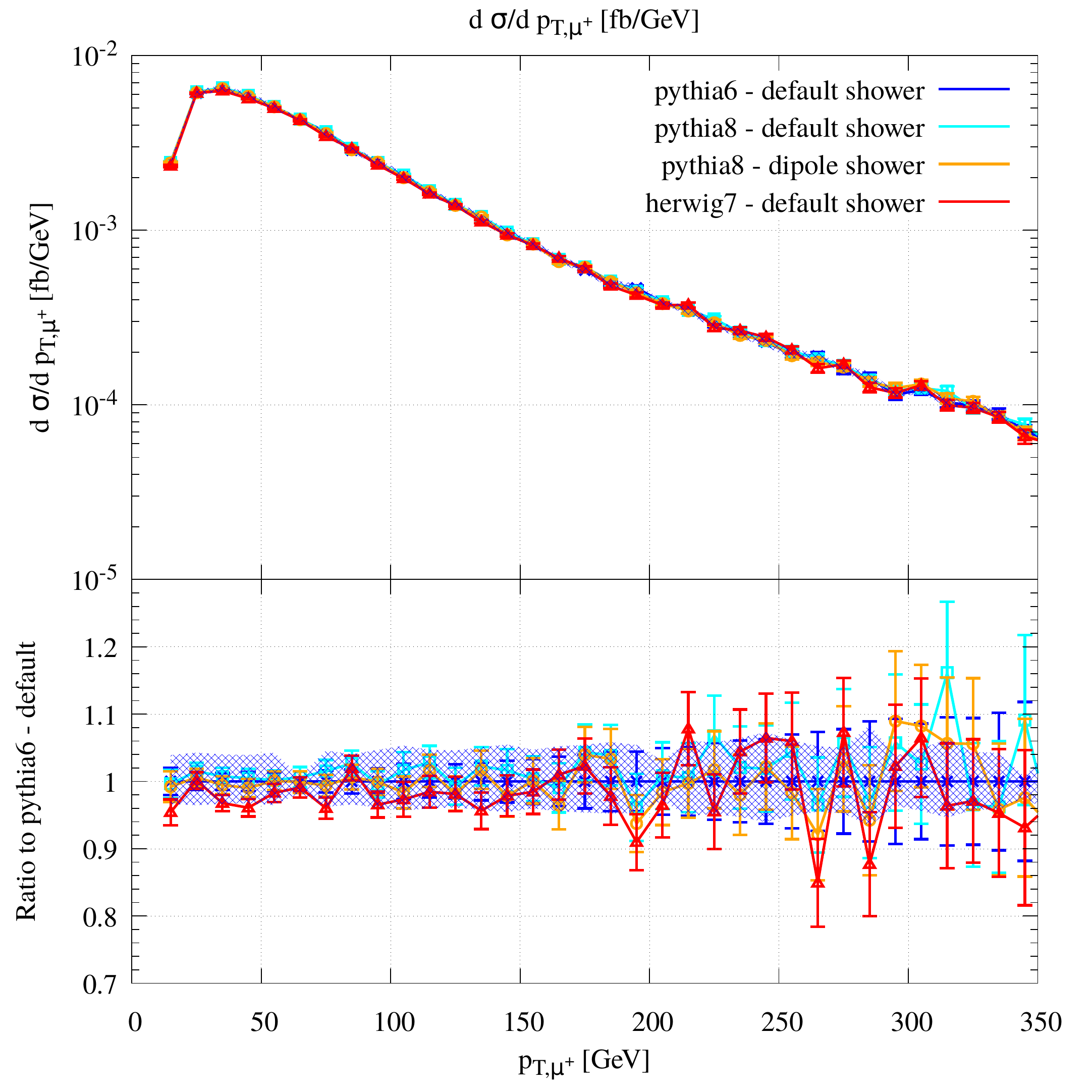}
\caption{
Transverse-momentum distributions of the hardest tagging jet (left) and of the positively charged muon (right) in the \ATLL{} scenario. In the lower panels the ratio of the respective distribution to the  \PYTHIAS{} reference result is shown. In each case the blue bands indicate the scale uncertainty of the \PYTHIAS{} simulations, statistical uncertainties are denoted by error bars. 
\label{fig:ptj1-ptmu1-atlas}
}
\end{figure*}

\begin{figure*}[tp]
\center
\includegraphics[width=0.49\textwidth]{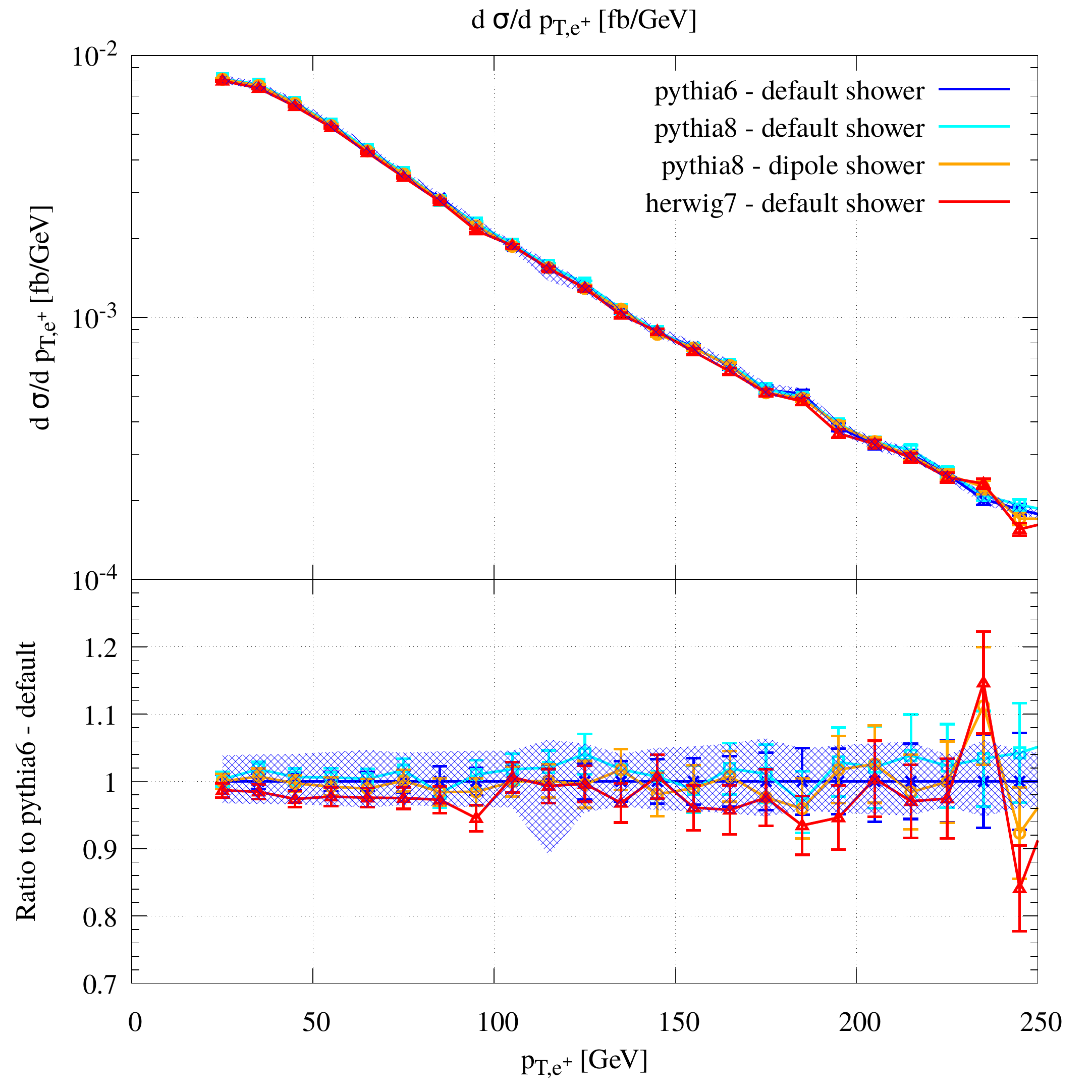}
\includegraphics[width=0.49\textwidth]{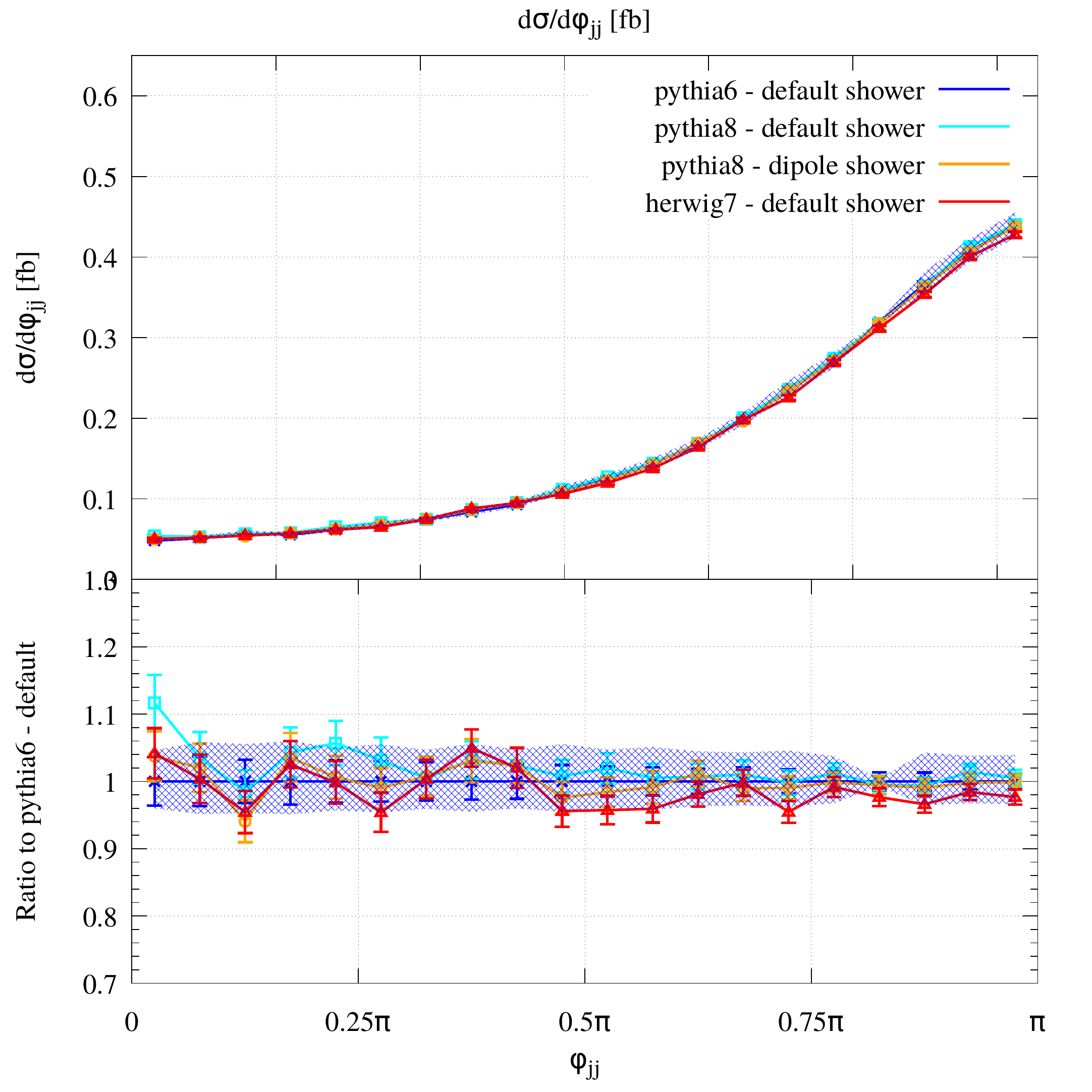}
\caption{
Transverse-momentum distribution of the hardest positron (left) and azimuthal-angle separation of the two tagging jets (right) in the \ATLL{} scenario. In the lower panels the ratio of the respective distribution to the  \PYTHIAS{} reference result is shown. In each case the blue bands indicate the scale uncertainty of the \PYTHIAS{} simulations, statistical uncertainties are denoted by error bars.  
\label{fig:pte1-phijj-atlas}
}
\end{figure*}

Figure~\ref{fig:ptj1-ptmu1-atlas} 
%
shows the transverse-momentum distributions of the hardest tagging jet and of the positively charged muon, while in fig.~\ref{fig:pte1-phijj-atlas} 
%
%
the transverse-momentum distribution of the hardest positron and the azimuthal-angle separation of the two tagging jets are illustrated for the \ATLL{} scenario. Analogous to the case of the \CMST{} setup, parton-shower dependencies are very small for distributions that are related to the two tagging jets. 

\begin{figure*}[tp]
\center
\includegraphics[width=0.49\textwidth]{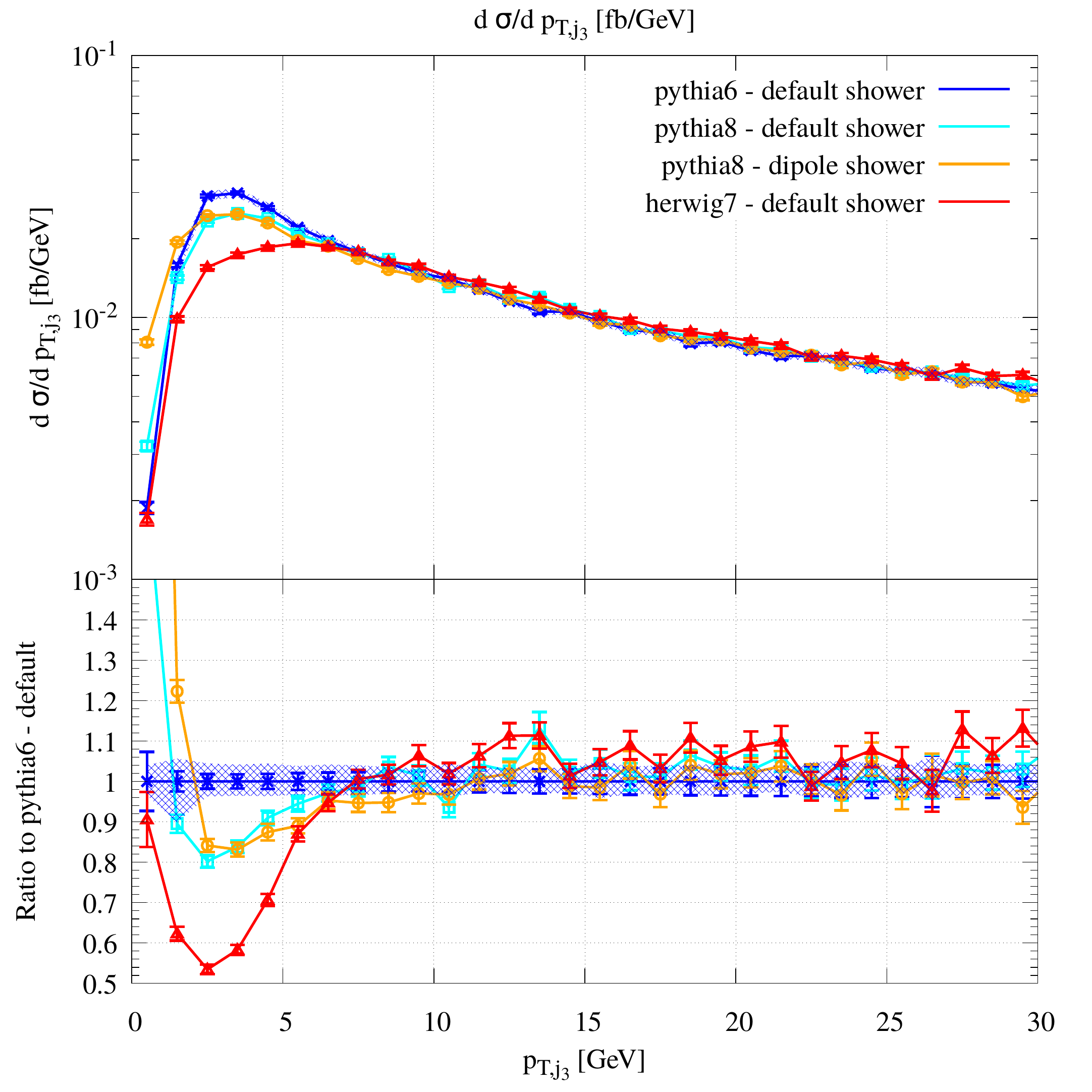}
\includegraphics[width=0.49\textwidth]{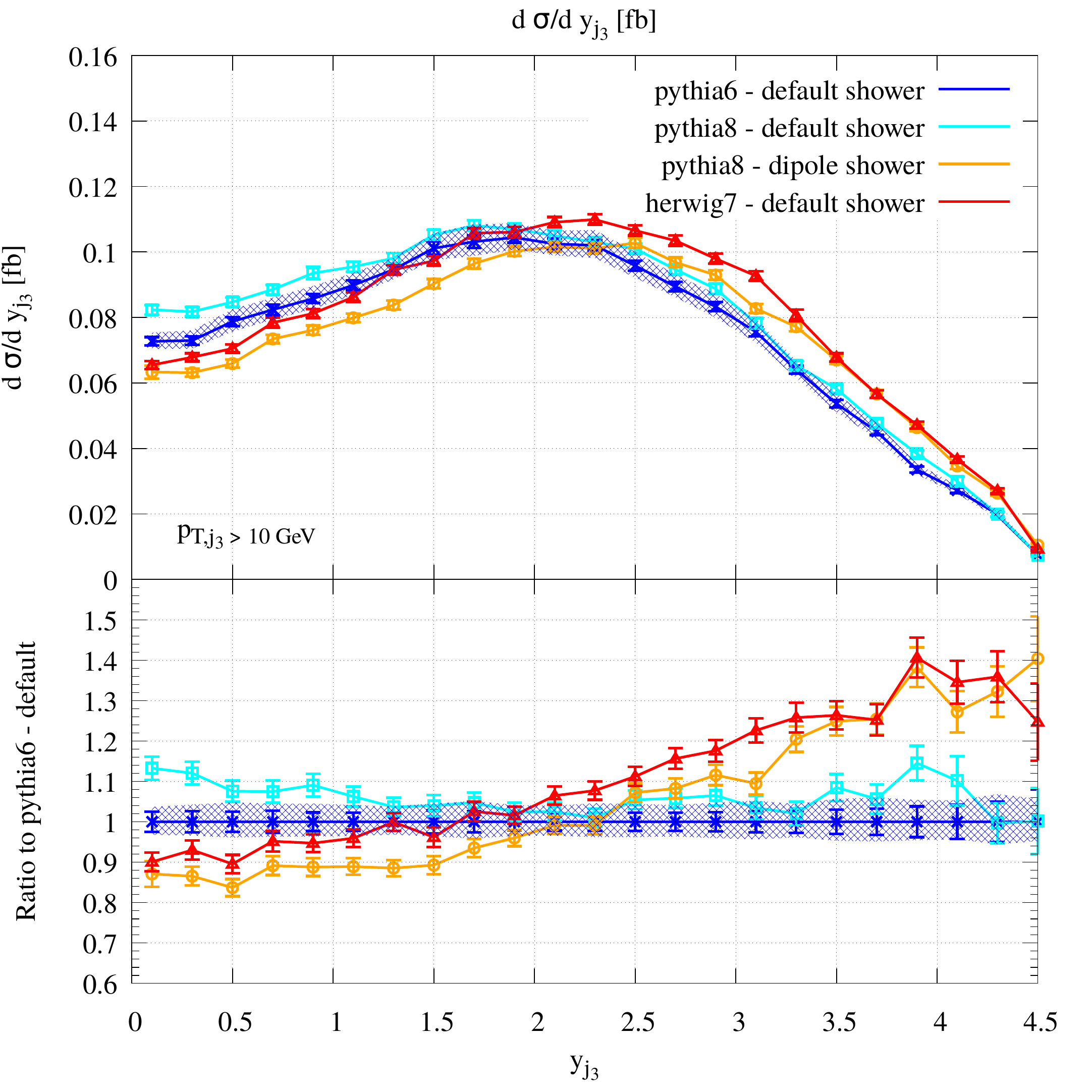}
\caption{ 
Transverse-momentum (left) and rapidity (right) distributions of the third jet in the \ATLL{} scenario. In the lower panels the ratio of the respective distribution to the  \PYTHIAS{} reference result is shown. In each case the blue bands indicate the scale uncertainty of the \PYTHIAS{} simulations, statistical uncertainties are denoted by error bars. 
\label{fig:ptj3-yj3-atlas}
}
\end{figure*}

\begin{figure*}[t]
	\centering
	\includegraphics[width=0.49\textwidth]{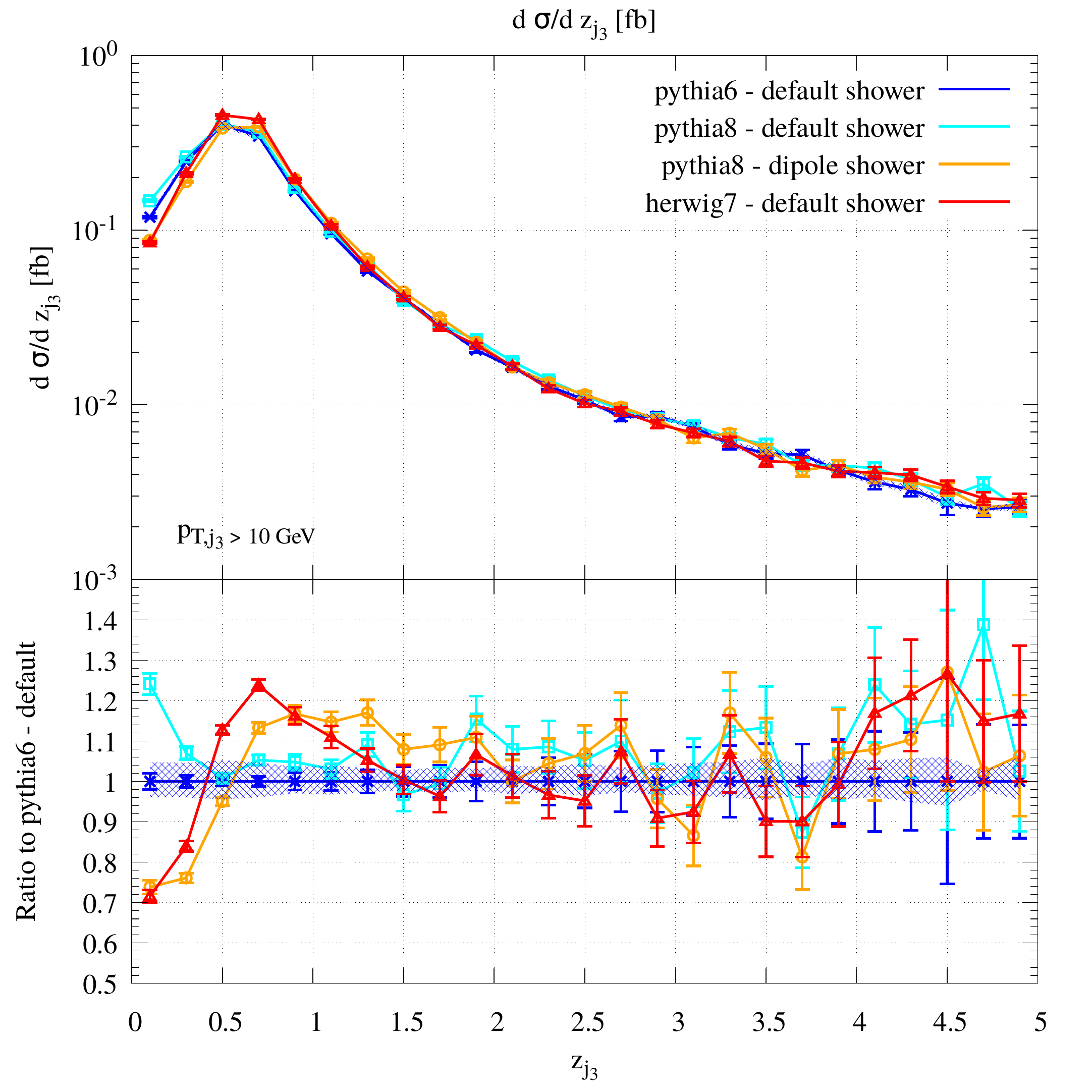}
	\includegraphics[width=0.49\textwidth]{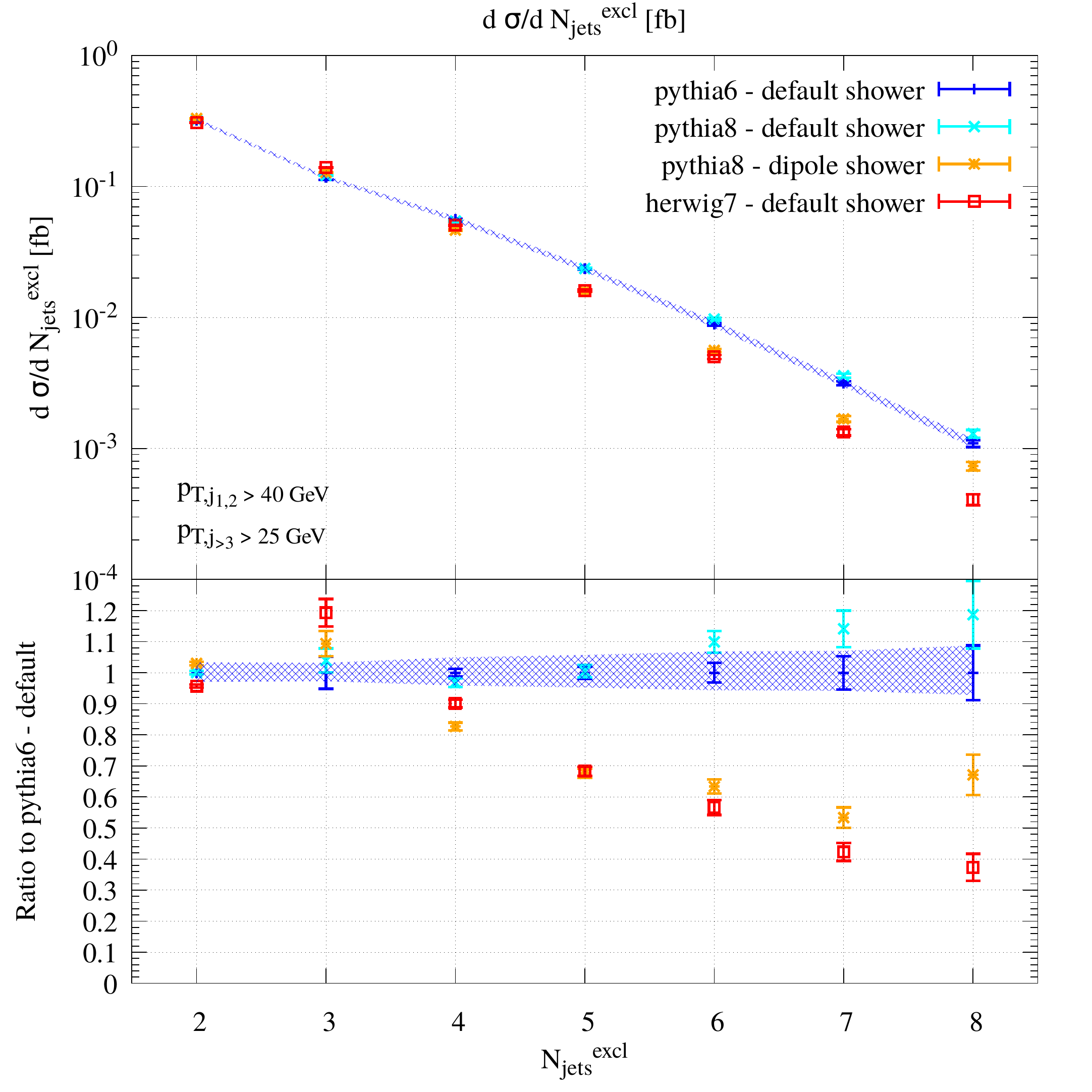}
	\caption{Differential distribution of the Zeppenfeld variable defined in eq.~\eqref{Zeppenfeld} (left) and the number of exclusive jets (right) in the \ATLL{} scenario. In the left plot we require $p_{T,j_3} > 10$ GeV and in the right plot we require all jets in addition to the two tag jets to have $p_T > 25$ GeV. In the lower panel the ratio of the respective distribution to the  \PYTHIAS{} reference result is shown. In each case the blue bands indicate the scale uncertainty of the \PYTHIAS{} simulations, statistical uncertainties are denoted by error bars.
          \label{fig:PSZj3-atlas}}
\end{figure*}

Larger differences between the various SMCs are again observed for distributions involving a non-tagging jet, such as the transverse-momentum and rapidity distributions of the third jet, shown in fig.~\ref{fig:ptj3-yj3-atlas},
%
%
and the Zeppenfeld variable and $N_{\mathrm{jets}}^{\mathrm{excl}}$ depicted in fig.~\ref{fig:PSZj3-atlas}. 
%
%
%
We note that, in general, differences between predictions obtained with various SMCs are slightly smaller in the \ATLL{} than in the \CMST{} setup, an effect we attribute to the more exclusive cut setup in the latter.  

\begin{figure*}[tp]
	\centering
	\includegraphics[width=0.49\textwidth]{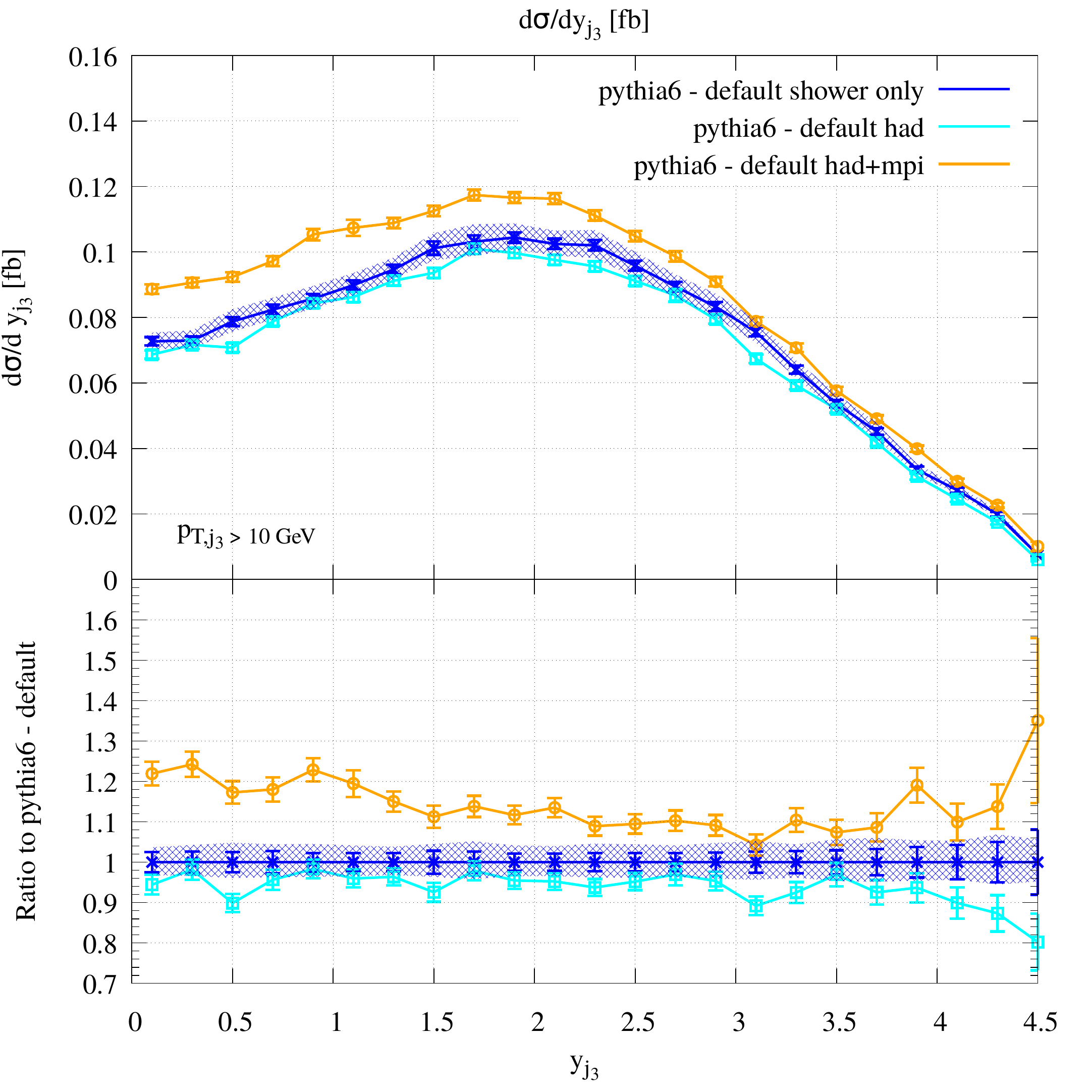}
	\includegraphics[width=0.49\textwidth]{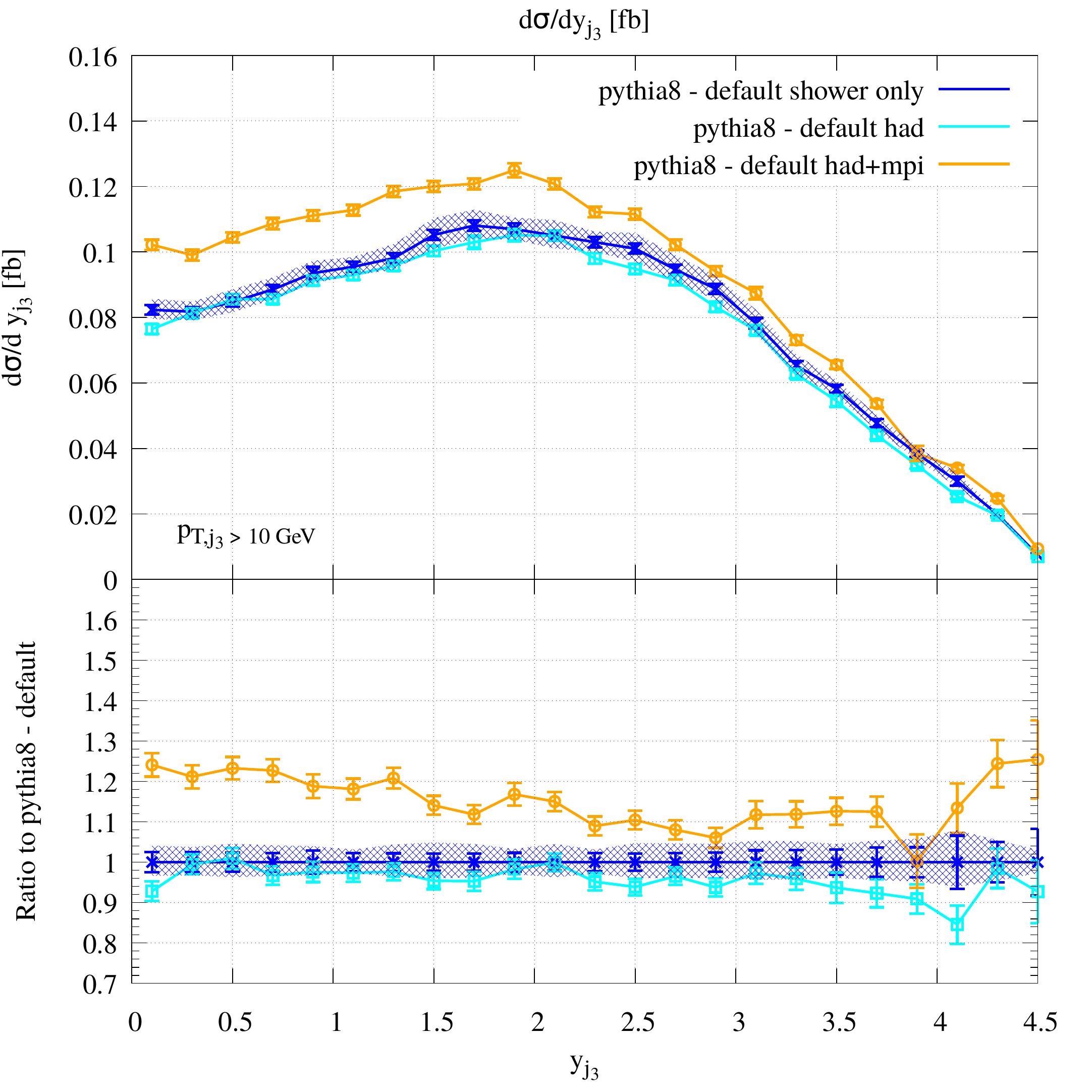}
	\includegraphics[width=0.49\textwidth]{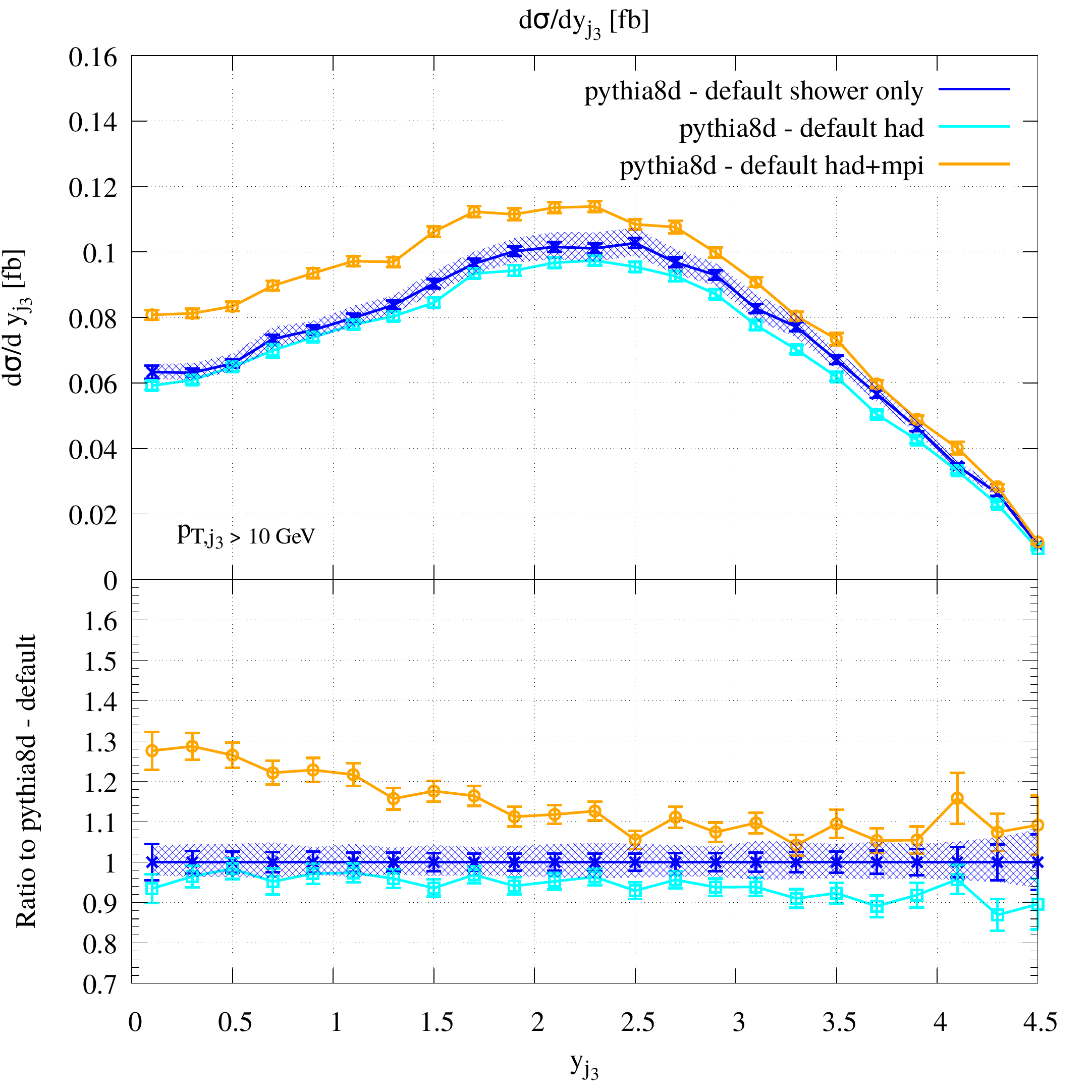}
	\includegraphics[width=0.49\textwidth]{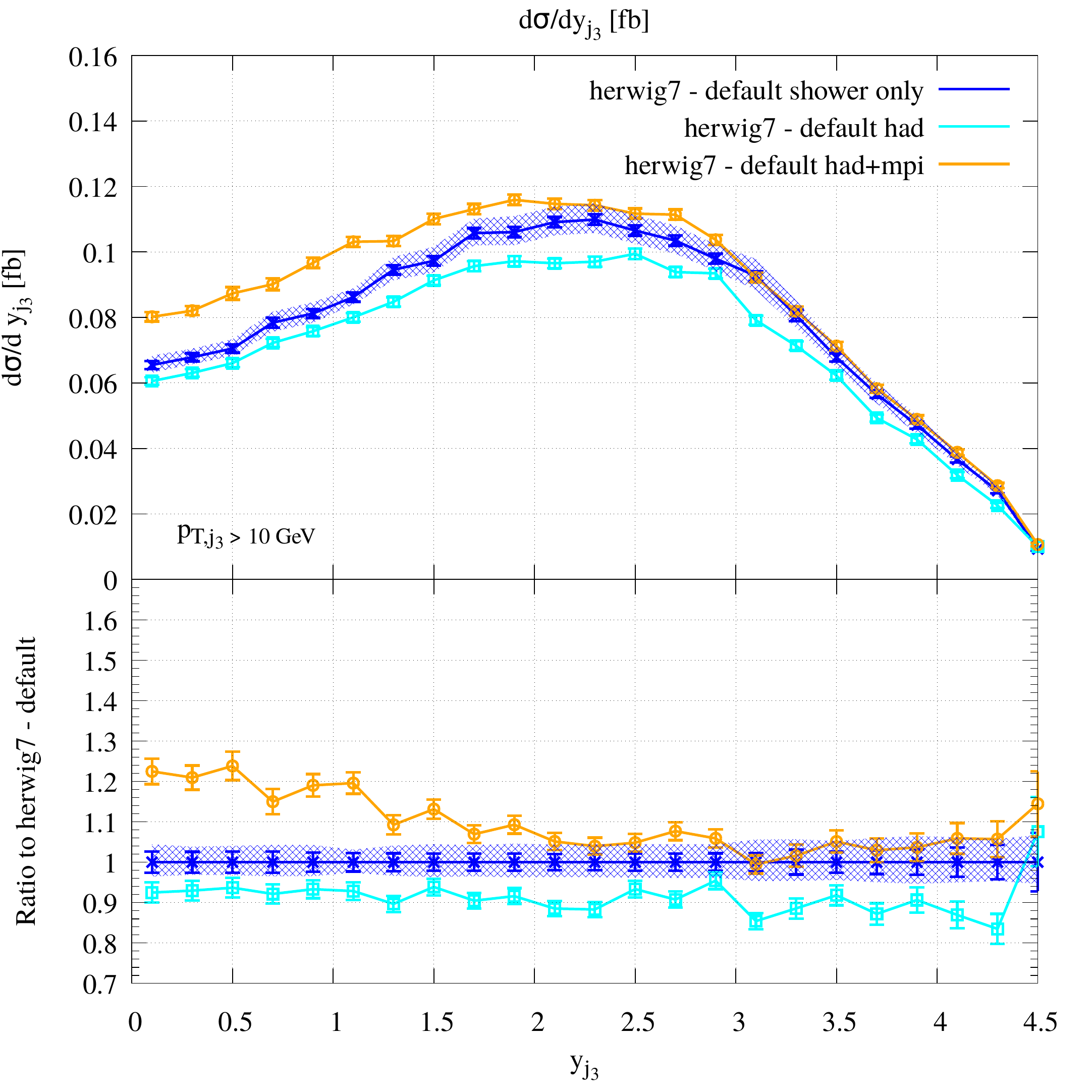}
\caption{Rapidity distribution of the third jet in the \ATLL{} scenario for different SMC settings [red: parton shower only, light blue: parton shower with hadronization, dark blue: parton shower with hadronization and multi-parton interactions]. The same observable is shown for \texttt{PYTHIA6} (a), \texttt{PYTHIA8} with recoil shower (b), \texttt{PYTHIA8} with dipole shower (c), and \texttt{HERWIG7} (d). 
 For each SMC, the lower panel shows the ratio $R_\mr{SMC}=d\sigma(\mr{SMC_{option}})/d\sigma(\mr{SMC_{default}})$ of the respective distribution. The blue bands indicate the scale uncertainty of the respective simulations without hadronization and MPI, statistical uncertainties are denoted by error bars. 
\label{fig:all-yj3-atlas}}
\end{figure*}

In fig.~\ref{fig:all-yj3-atlas} 
%
%
%
the rapidity distribution of the third jet is illustrated for different settings in the various SMCs to explore the impact of hadronization and multi-parton interactions in our predictions. In particular the impact of MPI is smaller here than in the case of the \CMST{} scenario considered in sec.~\ref{sec:pheno}.


\end{document}